\documentclass[oldversion]{aa}  
\usepackage{graphicx,float}
\usepackage{latexsym,amsmath,amssymb}
\usepackage{natbib}
\usepackage{changebar}
\usepackage{subfigure}
\usepackage{txfonts}
\usepackage{rotating}
\usepackage{longtable, lscape}

\usepackage{natbib}
\bibpunct{(}{)}{;}{a}{}{,}

\def\apjl{ApJ}%
\def\apjs{ApJS}%
\def\ao{Appl.~Opt.}%
\def\aap{A\&A}%
\begin{document}

   \title{The narrow Fe\,K$\alpha$ line and the molecular torus in active galactic nuclei - an IR/X-ray view}

   \subtitle{}

   \author{C. Ricci\inst{1,2}, Y. Ueda\inst{2}, K. Ichikawa\inst{2}, S. Paltani\inst{1}, R. Boissay\inst{1}, P. Gandhi\inst{3}, M. Stalevski\inst{4,5,6,7}, and H. Awaki\inst{8}
          }

   \institute{ Department of Astronomy, University of Geneva, ch. d'Ecogia 16, 1290 Versoix, Switzerland 
    \and Department of Astronomy, Kyoto University, Oiwake-cho, Sakyo-ku, Kyoto 606-8502, Japan
    \and Department of Physics, University of Durham, South Road, Durham DH1 3LE, UK 
    \and Astronomical Observatory, Volgina 7, 11060 Belgrade, Serbia 
    \and  Isaac Newton Institute of Chile, Yugoslavia Branch, Volgina 7, 11060 Belgrade, Serbia
    \and Sterrenkundig Observatorium, Universiteit Gent, Krijgslaan 281-S9, Gent, 9000, Belgium
    \and Departamento de Astronom'a, Universidad de Chile, Casilla 36-D, Correo Central, Santiago, Chile
    \and Department of Physics, Ehime University, Matsuyama, 790-8577, Japan \\
             }
    \offprints{ricci@kusastro.kyoto-u.ac.jp} 
   \authorrunning{ C. Ricci et al.}
   \titlerunning{Iron K$\alpha$ line and the molecular torus in AGN - an IR/X-ray view }
    \date{Received; accepted}

 
 \abstract{ The narrow component of the iron K$\alpha$ is an almost ubiquitous feature in the X-ray spectra of active galactic nuclei (AGN) and is believed to originate in neutral material, possibly located in the molecular torus. This would imply a tight connection between the Fe K$\alpha$ equivalent width (EW) and the physical properties of the torus. In a recent work we have shown that the decrease of the covering factor of the torus with the luminosity, as expected by luminosity-dependent unification models, would be able to explain the decrease of Fe K$\alpha$ EW with the luminosity (i.e., the X-ray Baldwin effect). Recent developments in the study of the mid-IR (MIR) spectrum of AGN allow important parameters of the torus to be deduced, such as its covering factor ($f_{\rm\,obs}$) and equatorial column density ($N_{\rm\,H}^{\rm\,T}$), by applying clumpy torus models. Using {\it XMM-Newton}/EPIC observations of a sample of 24 type-I AGN, we investigate the relation between the physical parameters of the torus obtained by recent MIR works and the properties of the Fe\,K$\alpha$ line. We correct the values of the Fe K$\alpha$ EW by taking the inclination angle, the photon index, the equatorial column density, and half-opening angle of the torus into account using a physical torus model of X-ray reprocessed radiation. We find that the relation between Fe\,K$\alpha$ EW and $f_{\rm\,obs}$ shows a slope that is consistent with the expected value, albeit with a low statistical significance. A trend that is consistent with the theoretical prediction is also found when comparing the Fe\,K$\alpha$ EW to $N_{\rm\,H}^{\rm\,T}$. Our work seems to confirm that the bulk of the narrow Fe\,K$\alpha$ line is produced by the same material responsible for the MIR emission.

 }
   \keywords{Galaxies: Seyferts -- X-rays: galaxies -- Galaxies: active -- Galaxies: nuclei 
                  }

   \maketitle

\section{Introduction}
The unification model of active galactic nuclei (AGN) predicts that the supermassive black hole (SMBH) in their centre is surrounded by a molecular toroidal-like structure \citep{Antonucci:1993kb}. Anisotropic obscuration was originally required to explain the detection of broad lines in polarised light found in the optical spectrum of the Seyfert\,2 \object{NGC\,1068} \citep{Antonucci:1985qo,Miller:1983lr}, and is now considered to be one of the fundamental ingredients needed to explain the structure of AGN. According to this paradigm Seyfert\,1s (Sy1s) are observed pole-on with respect to the molecular torus, while Seyfert\,2s (Sy2s) are seen edge-on. The radiation produced by the central engine and absorbed by the torus is mainly re-emitted in the mid-IR band (MIR, 5--30\,$\mu$m). The first direct observation of the dusty torus was carried out using MIR interferometry, for \object{NGC\,1068} \citep{Jaffe:2004fr}. This work was then followed by several others (e.g., \citealp{Prieto:2004uq}, \citealp{Prieto:2005kx}, \citealp{Meisenheimer:2007fk}, \citealp{Tristram:2007ys}, \citealp{Raban:2009vn}), and all of them detected a clear compact structure within few parsecs from the SMBH. 

The fluorescent iron K$\alpha$ line is possibly the most important tracer of the material surrounding the SMBH. The Fe\,K$\alpha$ line is made of two components, K$\alpha_1$ (E=6.404\,keV) and K$\alpha_2$ (E=6.391\,keV), with a branching ratio of K$\alpha_1$:K$\alpha_2$=2:1, and it is produced when one of the two K-shell electrons of an iron atom is ejected following photoelectric absorption of an X-ray photon. After the photoelectric event, the excited state can decay in two ways. i) An L-shell electron drops into the K-shell releasing a photon. ii) The excess energy is carried away through the ejection of an L-shell electron (Auger effect). The fluorescent yield ($Y$) determines the probability of fluorescence versus the Auger effect. The iron line is the strongest X-ray line produced from the reprocessing of the primary continuum, because of the Fe relative abundance, and because the fluorescent yield is proportional to the fourth power of atomic number ($Y\propto Z^4$). 

Amongst the other lines produced by X-ray reflection from neutral material the strongest are the iron K$\beta$ line at 7.06\,keV ($\sim13.5\%$ of the flux of the Fe\,K$\alpha$, \citealp{Palmeri:2003qf}), and the nickel K$\alpha$ line at $\sim7.47$\,keV (e.g., \citealp{Yaqoob:2011kx}). The first evidence of an Fe\,K$\alpha$ line in the X-ray spectrum of an AGN was found by \citet{Mushotzky:1978kx} when studying {\it OSO-8} observations of Centaurus\,A. Ten years later \citet{Guilbert:1988oq} and \citet{Lightman:1988qy} predicted that fluorescent emission from neutral iron should be common in the X-ray spectra of Seyfert galaxies. Following this, \citet{Nandra:1989uq} found evidence of an emission line at $E\sim6$\,keV in the {\it EXOSAT} spectrum of \object{MCG$-6-30-15$}. In the same year, using a larger sample \citet{Pounds:1989fj} found significant iron K$\alpha$ emission lines in the spectra of three more Seyfert\,1 galaxies: \object{NGC\,5548},  \object{NGC\,5506}, and  \object{NGC\,3227}. Since then, thanks to the enormous progress in the development of X-ray detectors, iron lines have been found to be almost ubiquitous in AGN (e.g., \citealp{Fukazawa:2011fk}).

The Fe\,K$\alpha$ line is made of two components. While the narrow core of the line, with a full width at half maximum (FWHM) of $\simeq 2,000\rm\,km\,s^{-1}$ \citep{Shu:2011fk}, is observed in almost all AGN, in $\simeq 35-45\%$ of the cases \citep{de-La-Calle-Perez:2010fk}, an additional broadened component due to relativistic effects (e.g., \citealp{Fabian:2003dp}) or to distortion of the continuum caused by clumpy ionised absorbers in the line-of-sight (e.g., \citealp{Turner:2009fk}, \citealp{Miyakawa:2012vn}) is found. The size of the narrow Fe\,K$\alpha$ emitting region is on average $\sim 3$ times larger than that of the broad line region \citep{Shu:2011fk}, which seems to point towards most of the narrow core originating in the molecular torus. Another argument in favour of this scenario is the weak variability of  reflection-dominated Compton-thick (CT, $N_{\rm\,H}\geq 10^{24}\rm\,cm^{-2}$) AGN (e.g., \citealp{Bianchi:2012dq} and references therein). A torus origin of the Fe\,K$\alpha$ line would imply that its equivalent width (EW) is directly linked to the half-opening angle of the torus $\theta_{\rm\,OA}$ \citep{Krolik:1994fk} and to its equatorial column density $N_{\mathrm{\,H}}^{\mathrm{\,T}}$ (e.g., \citealp{Ikeda:2009nx}, \citealp{Murphy:2009uq}).

An anti-correlation between the equivalent width of the Fe\,K$\alpha$ line and the X-ray luminosity of AGN has been found by a large number of studies (e.g., \citealp{Iwasawa:1993ys}, \citealp{Bianchi:2007vn}, \citealp{Shu:2010zr}). Such a trend is known as the X-ray Baldwin effect, for analogy with the Baldwin effect \citep{Baldwin:1977fk}, i.e. the decrease of the C\,IV$\,\lambda 1549$ EW with the luminosity. Several explanations have been proposed for the X-ray Baldwin effect: i) a luminosity-dependent variation in the ionisation state of the iron-emitting material \citep{Nandra:1997fk,Nayakshin:2000uq}; ii) the effect of the delay between the primary X-ray emission and the reflection component \citep{Jiang:2006vn,Shu:2012fk}; iii) the decrease in the number of continuum photons in the iron line region with the Eddington ratio ($\lambda_{\mathrm{Edd}}$, \citealp{Ricci:2013vn}), as an effect of the correlation between the photon index ($\Gamma$) of the continuum and $\lambda_{\mathrm{Edd}}$ (e.g., \citealp{Shemmer:2008fk}); iv) the decrease of the covering factor of the torus with the luminosity (e.g., \citealp{Page:2004kx}, \citealp{Zhou:2005ys}), as expected by luminosity-dependent unification models (e.g., \citealp{Ueda:2003qf}). In a recent paper \citep{Ricci:2013fk}, we have shown that the decrease of the covering factor of the torus with the luminosity is able to reproduce the slope of the X-ray Baldwin effect for a wide range of equatorial column densities of the torus.

The thermal MIR continuum is produced by circumnuclear dust (e.g., \citealp{Stalevski:2012fk}), which is heated by the optical/UV/X-ray photons produced in the disk and in the warm corona. The first attempts to model the MIR spectral energy distribution (SED) of AGN using a torus with a smooth density distribution (e.g., \citealp{Pier:1992ys,Pier:1993zr}) were able to model part of the SED but not to produce realistic MIR spectra. This was probably the first evidence that dust in the torus does not have a smooth distribution. Already years earlier, \citet{Krolik:1988ly} hypothesised that the torus is made of optically thick dusty clouds, because a smooth distribution could not survive close to the SMBH. Several other pieces of evidence of a clumpy structure of the torus have been added in past years. From interferometric observations of Circinus galaxy \citet{Tristram:2007ys} found that the data could not support a smooth-distribution scenario, but rather pointed towards the dust having a clumpy or filamentary structure. The discovery that Seyfert\,1s and Seyfert\,2s follow the same X-ray/MIR luminosity correlation (e.g., \citealp{Gandhi:2009uq}, \citealp{Ichikawa:2012fk}), and the detection of silicate emission in Seyfert\,2s \citep{Sturm:2006qf} also provide strong arguments for the clumpy scenario. Important information about the structure of the torus (e.g., its covering factor and number of clouds, see Section\,\ref{Sect:midIR_prop}) can be obtained by modelling the IR spectra of AGN using clumpy torus models such as the one developed by \citet{Nenkova:2002fk}. In the past few years, several studies (e.g., \citealp{Mor:2009fk}, \citealp{Alonso-Herrero:2011zr}) have carried out detailed analyses of the MIR properties of AGN for a significant number of objects. 

The aim of this work is to compare the properties of the narrow component of the iron K$\alpha$ line with those of the torus obtained by recent MIR studies.
The paper is organised as follows. In Sect.\,\ref{Sect:midIR_prop}, we present our MIR/X-ray AGN sample and describe the {\it XMM-Newton}/EPIC data analysis; in Sect.\,\ref{Sect:spectral_analysis} we illustrate the X-ray spectral analysis; in Sect.\,\ref{Sect:renorm} we discuss how to remove the effects of degeneracy caused by the X-ray photon index, the observing angle, the half-opening angle of the torus, and the torus equatorial column density on the values of Fe\,K$\alpha$ EW; and in Sect.\,\ref{Sect:EWvsf2} we study the relation between the Fe\,K$\alpha$ EW and the physical characteristics of the torus. In Sect.\,\ref{Sect:summary} we discuss our findings and present our conclusions.

\begin{table*}
\begin{center}
\caption[]{{\it XMM-Newton} observation log.}
\resizebox{0.93\textwidth}{!}{
\label{tab:obslog}
\begin{tabular}{llcccccc}
\hline \hline \noalign{\smallskip}
 & (1) & (2) &  (3) & (4) & (5) & (6) & (7) \\
\noalign{\smallskip}
 Source & N &$N_{\rm\,H}^{\rm\,G}$ &  $z$ & Obs. mode/filter & Obs. date & Obs. ID & Net exposure\\
\noalign{\smallskip}
  & & [{\tiny $10^{20}\rm\,cm^{-2}$}]  & & & {\tiny YYYY-MM-DD} & & [{\tiny ks}] \\
\noalign{\smallskip}
\hline \noalign{\smallskip}
\object{B2 2201+31A}  & &11.8 & 0.2980 & FF/M; FF/M; FF/M & 2008-12-01 &0550871001 & 13.5/13.2/13.2 \\
\noalign{\smallskip}
\object{IC 4329A} & I &4.42  &	0.0160		 & FF/M$^{\mathrm{P}}$; LW/M$^{\mathrm{P}}$; LW/M$^{\mathrm{P}}$ & 2001-01-31 & 0101040401 & 10.4/10.4/10.4 \\
		 & II &  & &  SW/T; LW/M; LW/T & 2003-08-06 &  0147440101  &  118.4/118.3/118.3  \\
\noalign{\smallskip}
\object{NGC 3227} & I &2.13  &	0.0037		 & FF/M; FF/M; FF/M & 2000-11-29 & 0101040301 & 34.6/34.4/34.4 \\
	& II & &		&	LW/M; SW/M; SW/M		&	2006-12-03 	&	0400270101	&	101.3/101.0/101.0		\\
\noalign{\smallskip}
\object{NGC 4151}  & I &1.99  &	0.0033			&	FF/M; FF/M$^{\mathrm{P}}$; FF/M$^{\mathrm{P}}$			&	2000-12-22	&	0112830201	&	57.0/56.9/56.9			\\
	& II &&	 & SW/M; SW/M; SW/M & 2000-12-22 & 0112310101 & 29.9/29.3/29.3 \\
	&  III &&			 & FF/M; FF/M; FF/M & 2000-12-22 & 0112830501 & 19.7/19.6/19/6 \\
	& IV  &&			 & SW/M; SW/M$^{\mathrm{P}}$ ; SW/M$^{\mathrm{P}}$   & 2003-05-25 & 0143500101& 18.5/18.3/18.4 \\
	& V  &&			 & SW/M; SW/M$^{\mathrm{P}}$; SW/M$^{\mathrm{P}}$ & 2003-05-27 & 0143500201 & 18.4/18.1/18.1 \\
	& VI  &&			 & SW/M; SW/M$^{\mathrm{P}}$; SW/M$^{\mathrm{P}}$  & 2003-05-27 & 0143500301 & 18.5/18.0/18.0 \\
	& VII  &&			 & SW/M; SW/M; SW/M & 2006-05-16 & 0402660101 & 40.0/39.8/39.8 \\
	& VIII  &&			 & SW/M; SW/M; SW/M  & 2006-11-30 & 0402660201 & 46.5/37.5/37.4 \\
\noalign{\smallskip}
\object{NGC 6814} &  &  12.8 &	 0.0052		&	FF/M$^{\mathrm{P}}$; FF/M$^{\mathrm{P}}$; FF/M$^{\mathrm{P}}$	&	2009-04-22	&	0550451801	&	28.4/28.4/28.4 			\\
\noalign{\smallskip}
\object{NGC 7469} & I &4.86  &	0.0159		 & SW/M; LW/M$^{\mathrm{P}}$; FF/M$^{\mathrm{P}}$   & 2000-12-26 & 0112170101 & 17.6/17.1/17.1 \\
			& II  &	&		 & SW/M; LW/M; FF/M & 2000-12-26 & 0112170301 & 23.1/22.6/22.6 \\
		& III &	&		&	SW/M; --; --	&	2004-11-30	&	0207090101	&	84.5/--/--	\\
		& IV  &  &			 & SW/M; --; -- & 2004-12-03 & 0207090201 & 78.6/--/-- \\
\noalign{\smallskip}
\object{PG 0050+124}   & I&5.00 & 0.0609 & LW/M$^{\mathrm{P}}$; SW/M$^{\mathrm{P}}$; SW/M$^{\mathrm{P}}$	&	2002-06-22	&	0110890301	&	19.3/19.2/19.2	\\
    			& II & &			  &  SW/M; SW/M; SW/M & 2005-07-18 & 0300470101 & 77.6/77.3/77.3 \\
\noalign{\smallskip}
\object{PG 0157+001}  & &2.59 & 0.1628 & FF/M; LW/M; LW/M & 2000-07-29 & 0101640201 & 7.5/7.5/7.5 \\
\noalign{\smallskip}
\object{PG 0838+770}   & &2.09 & 0.1313 &FF/M; FF/M; FF/M &2009-03-02  & 0550870401 & 17.6/17.5/17.5 \\
\noalign{\smallskip}
\object{PG 0953+414}  & &1.14 & 0.2390 & LW/T; LW/T; LW/T  & 2001-11-22 & 0111290201 & 11.3/11.3/11.3 \\
\noalign{\smallskip}
\object{PG 1004+130}   & &3.70 & 0.2406 & FF/M; FF/M; FF/M & 2003-05-04 & 0140550601 & 20.4/20.2/20.2 \\
\noalign{\smallskip}
\object{PG 1116+215}   & I &1.28 &0.1765  &  LW/T; LW/T; LW/T    & 2001-12-02 & 0111290401 & 7.0/6.0/6.0 \\
    		     	& II  & &			   &SW/T; SW/T; SW/T  & 2004-12-17 & 0201940101 & 98.1/86.3/86.3 \\
			& III  & & 			   & SW/T; SW/T; SW/T & 2004-12-19 & 0201940201 & 7.2/6.8/6.8 \\
			& IV  & &			   &  SW/T; SW/T; SW/T & 2008-05-27 & 0554380101 & 85.7/84.0/84.0 \\
			&  V & &			   & SW/T; SW/T; SW/T & 2008-12-15 & 0554380201 & 88.1/71.2/87.9 \\
			& VI  & &			   & SW/T; SW/T; SW/T  & 2008-12-20 & 0554380301 & 87.6/79.4/79.4 \\
\noalign{\smallskip}
\object{PG 1126$-$041}  & I & 4.30 &  0.0600 & LW/T; LW/T; LW/T & 2004-12-31 & 0202060201 & 31.3/31.0/31.0 \\
    			& II  & &		    	   &FF/M; FF/M; FF/M & 2008-06-15 & 0556230701 & 15.0/15.0/15.0 \\
			  & III  &	&		   & FF/M; FF/M; FF/M & 2008-12-13 & 0556231201 & 5.0/5.0/5.0 \\
			& IV  &		&	   & FF/M; FF/M; FF/M & 2009-06-21 & 0606150101 & 101.0/100.9/100.9 \\
\noalign{\smallskip}
\object{PG 1229+204}   & &2.21 & 0.0637 &SW/M; FF/M; LW/M &2005-07-09 & 0301450201 & 24.3/24.3/24.3 \\
\noalign{\smallskip}
\object{PG 1244+026}   & &1.75 & 0.0482 &FF/T; FF/M$^{\mathrm{P}}$; FF/M$^{\mathrm{P}}$ & 2001-06-17 & 0051760101 & 6.3/6.2/6.2 \\
\noalign{\smallskip}
\object{PG 1309+355}   & &1.03 & 0.1829  & LW/T; LW/T; LW/T & 2002-06-10  & 0109080201 & 25.3/25.3/25.3 \\
\noalign{\smallskip}
\object{PG 1411+442}  & &1.15 & 0.0896  & FF/T; FF/M; FF/M & 2002-07-10 &0103660101 & 25.0/24.8/24.8 \\
\noalign{\smallskip}
\object{PG 1426+015}   & &2.85 & 0.0860 & FF/T$^{\mathrm{P}}$; LW/T$^{\mathrm{P}}$; LW/TK$^{\mathrm{P}}$ & 2000-07-28 &0102040501& 6.0/4.0/4.0 \\ 
\noalign{\smallskip}
\object{PG 1435$-$067}  & &5.34 & 0.1290 & FF/M; FF/M; FF/M & 2009-02-02  & 0550870201 & 14.2/14.2/14.2 \\
\noalign{\smallskip}
\object{PG 1440+356}  & I&1.03 & 0.0770 & LW/T; LW/T$^{\mathrm{P}}$; LW/T$^{\mathrm{P}}$ & 2001-12-23 & 0107660201 & 26.2/26.0/26.0 \\
   			& II  &	&		& SW/M; SW/M; SW/M & 2003-01-02 & 0005010101 & 24.6/24.4/24.4 \\
			&  III &	 &		& SW/M; SW/M; SW/M & 2003-01-04 & 0005010201 & 27.5/27.0/27.0 \\
			&  IV &	  &		& SW/M; SW/M; SW/M & 2003-01-07 & 0005010301 & 26.0/25.8/25.8 \\
\noalign{\smallskip}
\object{PG 1448+273}   & &2.44 & 0.0645 & LW/M; LW/M; LW/M & 2003-02-08  & 0152660101 & 19.5/19.5/19.5 \\
\noalign{\smallskip}
\object{PG 1613+658}  & I&2.87 & 0.1385  & FF/T; LW/T; LW/TK & 2001-04-13 & 0102040601 &  5.3/1.0/1.0 \\
			& II  &		&	  & FF/T; LW/T; LW/TK & 2001-08-29 & 0102041301 & 2.7/2.7/2.7 \\
\noalign{\smallskip}
\object{PG 1626+554}   & &1.96 & 0.1320 & LW/T; LW/T; LW/T & 2002-05-05 &0109081101 & 6.1/5.4/5.4 \\
\noalign{\smallskip}
\object{PG 2214+139}   & &4.96 & 0.0663 & FF/T; FF/T; FF/T & 2002-05-12 & 0103660301 & 25.3/25.2/25.2  \\
\noalign{\smallskip}
\hline
\noalign{\smallskip}
\multicolumn{8}{l}{{\bf Note.} (1) the number of the observation, (2) the value of the Galactic column density ($N_{\rm\,H}^{\rm\,G}$) in the direction of the source } \\
\multicolumn{8}{l}{(from \citealp{Dickey:1990uq}), (3) the redshift ($z$), (4) the modes and filters of PN, MOS1 and MOS2 exposures,}\\
\multicolumn{8}{l}{(5) the date and (6) the ID of the observation,  and (7) the net exposure times. $^{\mathrm{P}}$ Piled-up exposure. }\\
\multicolumn{8}{l}{The modes and filters of EPIC are the following. FF: full frame; LW: large window;SW: small window; T: thin; M: medium; TK: thick.} \\
\end{tabular}
}
\end{center}
\end{table*}

\begin{figure}[t!]
\centering
5
\centering
\includegraphics[height=9cm,angle=270]{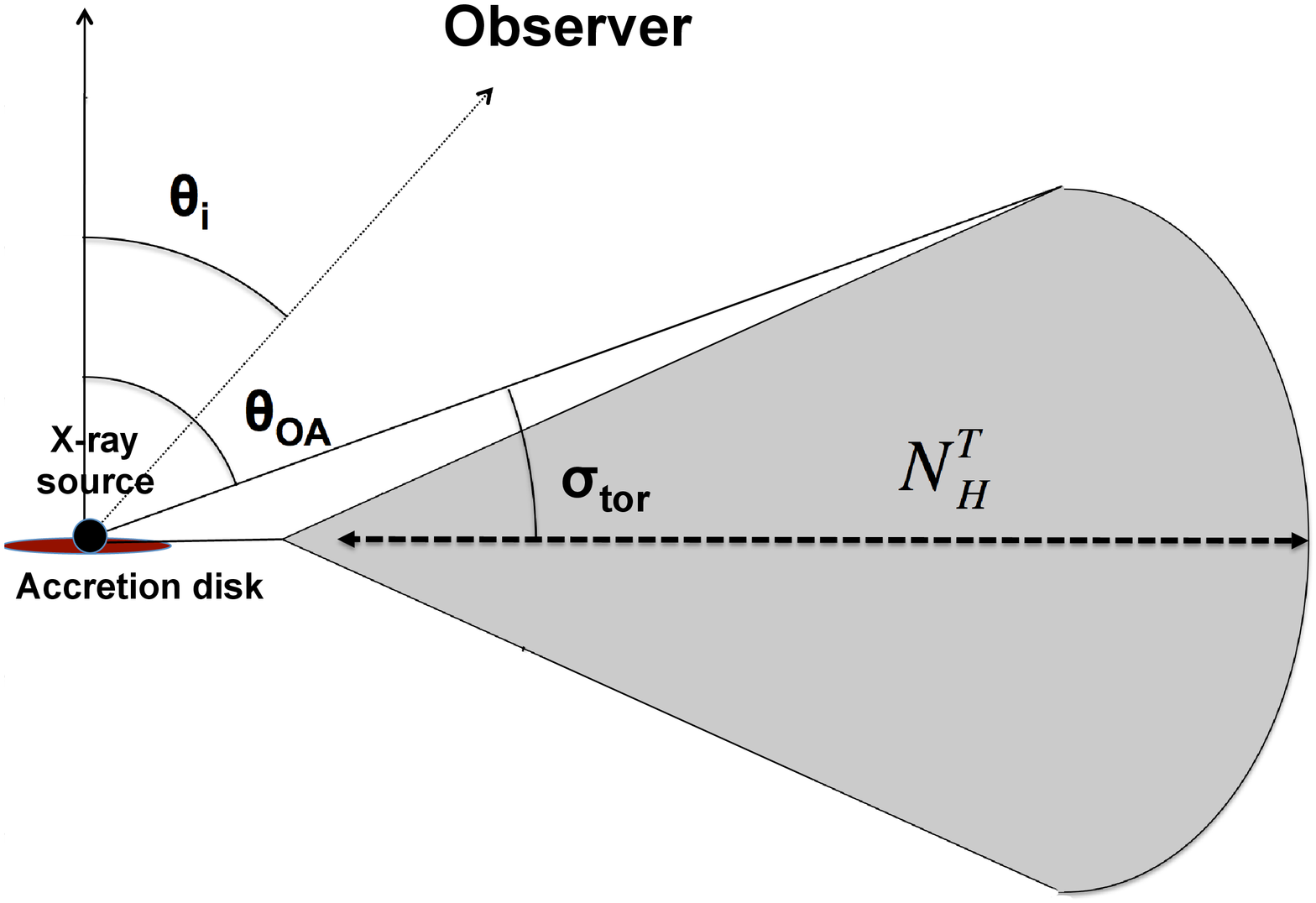}
  \caption{Schematic representation of the angles considered. $\theta_{\rm\,i}$ is the inclination angle of the observer, while $\theta_{\rm\,OA}$ ($\pi/2-\sigma_{\rm\,tor}$) is the half-opening angle of the torus. $N_{\rm\,H}^{\rm\,T}$ is the equatorial column density of the torus, i.e. the maximum value of $N_{\rm\,H}$ for any value of $\theta_{\rm\,i}$.}
\label{fig:geometry}

\end{figure}

\section{Sample and X-ray data analysis}\label{Sect:midIR_prop}

To study the relation between the iron K$\alpha$ line and the properties of the torus obtained from MIR studies, we used the sample reported in the recent work of \cite{Elitzur:2012vn}, which includes the works of \cite{Mor:2009fk}, \citet{Nikutta:2009kx}, \citet{Alonso-Herrero:2011zr}, \citet{Deo:2011vn}, and \citet{Ramos-Almeida:2011ly}. All these works used the IR clumpy torus model of \citet{Nenkova:2002fk,Nenkova:2008uq,Nenkova:2008kx}, which allows fundamental characteristics of the torus to be deduced by fitting MIR spectra. In the model, the optical luminosity is used to estimate the bolometric emission of the accreting system irradiating the torus. The free parameters of this model are the torus width parameter ($\sigma_{\mathrm{tor}}=\pi/2-\theta_{\rm\,OA}$, see Fig.\,\ref{fig:geometry}), the mean number of clouds along the equatorial line ($N_0$), the $5500\,\AA$ dust optical depth of a single cloud ($\tau_{\mathrm{V}}$), the inclination angle of the torus with respect to the line of sight ($\theta_{\mathrm{\,i}}$), the ratio between the outer and inner radius of the torus ($Y$), and the index of the radial power-law distribution of clouds (q, where the number of clouds is given by $N(r)\propto r^{\mathrm{-q}}$).

We cross-correlated the sample of \cite{Elitzur:2012vn} with the {\it XMM-Newton} \citep{Jansen:2001fk} public data archive (as of November 2012), selecting only type-I AGN to avoid uncertainties in the estimates of the Fe\,K$\alpha$ and continuum flux due to absorption. Amongst the sources with public {\it XMM-Newton} observations, PG\,1700+518 was detected with a very low S/N, which did not allow constraining the parameters of the Fe K$\alpha$ line, so that its spectrum was not used for our study. The final sample contains a total of 49 observations of 24 objects. Most of the sources (19) in our final sample are from the work of \citet{Mor:2009fk}, while four sources are taken from \citet{Alonso-Herrero:2011zr} and only one from \citet{Ramos-Almeida:2011ly}. None of the sources reported in the works of \citet{Nikutta:2009kx} and \citet{Deo:2011vn} were observed by {\it XMM-Newton}. Although all the works use the clumpy torus model of \citet{Nenkova:2002fk,Nenkova:2008uq,Nenkova:2008kx}, some differences exist in the approach they followed. \citet{Mor:2009fk} fitted the {\it Spitzer}/IRS $\sim2-35\,\mu$m spectra using a three-component model, which includes a dusty clumpy torus, a clumpy narrow-line region (NLR), and black-body emission from hot dust. This last component accounts for the near-IR (NIR, $\lambda\lesssim 5\mu$m) excess observed when fitting the spectra using only the first two components. In a recent work \citet{Mor:2011kx} studied a large sample of $\sim 15,000$ AGN and show that most AGN need this hot dust component to explain their NIR spectra. \citet{Alonso-Herrero:2011zr} combined the IR photometric SED with MIR ground-based spectroscopic data in the $8-13\,\mu$m, while \citet{Ramos-Almeida:2011ly} only used photometric data. Both \citet{Alonso-Herrero:2011zr} and \citet{Ramos-Almeida:2011ly} fitted the data using only the clumpy torus model, because the high angular resolution data they use in their work allows contamination from NLR dust to be ignored.

To study the Fe K$\alpha$ EW we used the data obtained by the PN \citep{Struder:2001uq} and MOS \citep{Turner:2001fk} cameras on-board {\it XMM-Newton}. The original data files (ODFs) were downloaded from the {\it XMM-Newton} Science Archive (XSA)\footnote{http://xmm.esac.esa.int/xsa/} and then reduced using the {\it XMM-Newton} Standard Analysis Software (SAS) version 12.0.1 \citep{Gabriel:2004fk}. The raw PN and MOS data files were processed using the \texttt{epchain} and \texttt{emchain} tasks, respectively. For each observation we checked the background light curve in the 10--12 keV energy band of the data sets in order to detect and filter the exposures for periods of high background activity. We selected only patterns that correspond to single and double events (PATTERN~$\leq 4$) for PN, and to single, double, triple, and quadruple events for MOS (PATTERN~$\leq 12$), as suggested by the standard guidelines. The source spectra were extracted from the final filtered event list using circular regions centred on the object (with a typical radius of 30\,arcsec), while the background was estimated from regions close to the source (preferably on the same CCD), where no other source was present (with a radius of 40\,arcsec). For sources detected with a low S/N, we extracted the spectra using a smaller radius (10\,arcsec). We checked for pile-up with the \texttt{epatplot} task, and for those observations where it was significant (see Table\,\ref{tab:obslog}), we used annular regions centred on the source, with an inner radius of 5 to 15\,arcsec, depending on the strength of the pile-up. We added a multiplicative factor to the models to account for cross-calibration between PN and MOS. We fixed the factor to 1 for EPIC/PN and left the MOS1 and MOS2 factors free. For all the spectra the value of the factor turned out to be close to one within a few percentage points.
The ancillary response matrices (ARFs) and the detector response matrices (RMFs) were generated using the tasks \texttt{arfgen} and \texttt{rmfgen}, respectively. The spectra were grouped to have at least 20 counts per bin, in order to use $\chi^{2}$ statistics.

The list of AGN used, together with the values of their redshift ($z$), of the Galactic column density in their direction ($N_{\rm\,H}^{\rm\,G}$), and their X-ray observation log is reported in Table\,\ref{tab:obslog}.

\begin{figure*}[t!]
\centering
\begin{minipage}[!b]{.48\textwidth}
\centering
\includegraphics[height=10cm,angle=270]{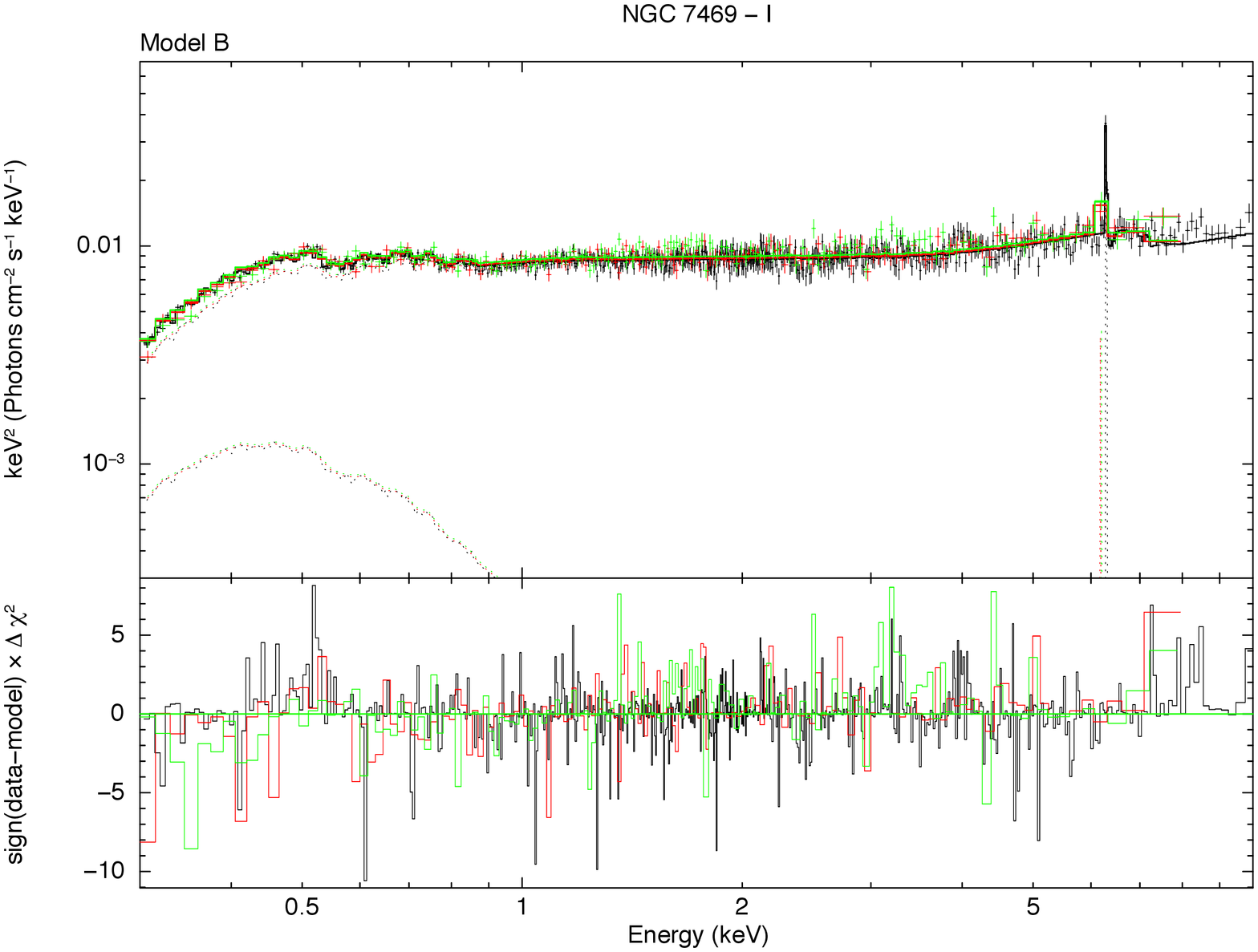}\end{minipage}
\hspace{0.05cm}
\begin{minipage}[!b]{.48\textwidth}
\centering
\includegraphics[height=11cm,angle=270]{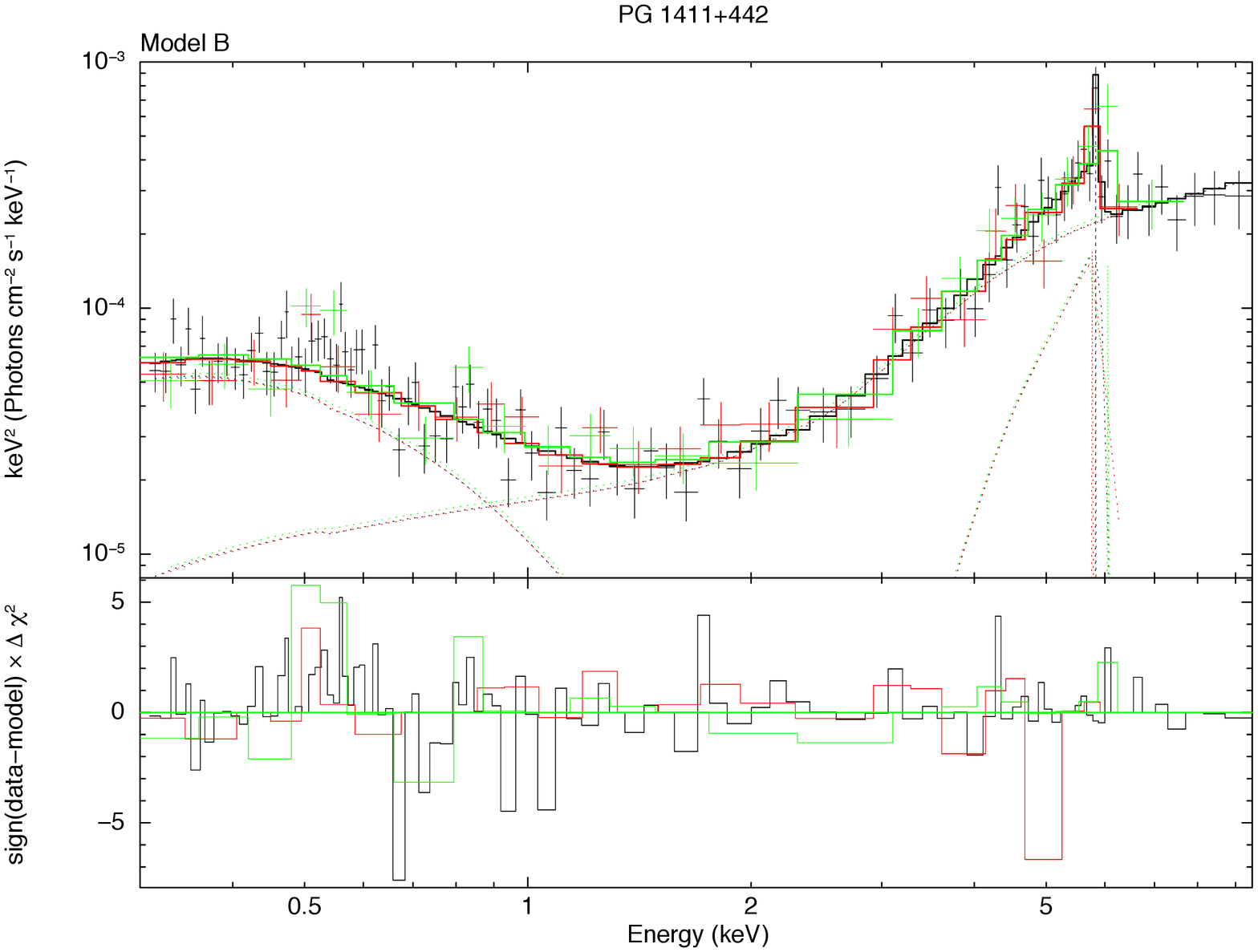}\end{minipage}
 \begin{minipage}[t]{1\textwidth}
  \caption{{\it XMM-Newton} EPIC spectra of the first observation of NGC\,7469 ({\it left panel}) and of PG\,1411+442 ({\it right panel}). The black points represent EPIC/PN data, and the red and green MOS data. The dotted lines show the different components of the models used. In the case of PG\,1411+442 besides a power law and a narrow Fe K$\alpha$ line, a broad relativistic line was added. Details on the models and on their parameters are reported in Appendices\,\ref{Appendix0} and \ref{Appendix1}. }
\label{fig:xrayspec}
 \end{minipage}
\end{figure*}

\section{X-ray spectral analysis}\label{Sect:spectral_analysis}
The X-ray spectral analysis was carried out using XSPEC\,12.7.1b \citep{Arnaud:1996kx}. Since we are dealing with objects that may have different characteristics in the X-rays, we started the analysis from a simple baseline model and then added absorbing or emitting components to improve the $\chi ^{2}$. More complex models were adopted based on the results of the F-test, using a probability of $p=95\%$ as a threshold.
The baseline model consists of a power-law continuum absorbed by Galactic absorption plus a Gaussian line to account for the iron K$\alpha$ emission (\texttt{wa$_{\mathrm{G}}$*(zpo+zgauss)} in XSPEC). 

The most common features observed in the X-ray spectra of type-I AGN are ionised absorbers (often called warm absorbers) and a soft excess. Ionised absorption is believed to be produced in disk outflows (e.g., \citealp{Turner:2009fk}) and was accounted for using the \texttt{zxipcf} model \citep{Reeves:2008kx}. This multiplicative model uses a grid of XSTAR \citep{Kallman:2001fk,Bautista:2001uq} photoionised absorption models, and its free parameters are the covering factor of the ionised absorber $f_{\rm\,W}$, its column density ($N_{\rm\,H}^{\rm\,W}$) and its ionisation parameter ($\xi$). The ionisation parameter is given by $\xi=L_{\mathrm{ion}}/nr^2$, where $L_{\mathrm{ion}}$ and $r$ are the 5\,eV-300\,keV luminosity and distance from the absorber of the ionising source, respectively, while $n$ is the density of the absorber. The origin of the soft excess is still controversial and might be related to blurred reflection (e.g., \citealp{Crummy:2006vn}), to Comptonisation of ultraviolet disk photons in a plasma cooler than the one responsible for the primary continuum (e.g., \citealp{Mehdipour:2011ys}, \citealp{Noda:2013fk}), or to smeared absorption (e.g., \citealp{Gierlinski:2004zr}). Since we are not interested in a detailed analysis of the soft excess, we adopted a simple phenomenological model (\texttt{bremsstrahlung}) to account for this feature.
All the sources of the sample require more complex models than the baseline. The models we applied to fit the X-ray spectra are the following (listed in the order in which they were applied):\smallskip\newline
{\bf Model A}. Baseline model and a bremsstrahlung component at low energies to represent the soft excess. In XSPEC this is written as \texttt{wa$_{\mathrm{G}}$*(zpo + bremss + zgauss)}. The free parameters of this model are the photon index of the power-law continuum ($\Gamma$), the temperature of the bremsstrahlung (kT), the energy of the Fe K$\alpha$ line ($E_{\rm\,K\alpha}$), and the normalisations of the three components. This model was used for 19\,observations.\smallskip\newline
{\bf Model B}.  Baseline model absorbed by a partially covering warm absorber: \texttt{wa$_{\mathrm{G}}$*zxipcf(zpo +zgauss)}. The free parameters are those of the baseline model, plus the parameters of the warm absorber ($\xi$, $N_{\rm\,H}^{\rm\,W}$ and $f_{\rm\,W}$). This model was adopted for three observations.\smallskip\newline
{\bf Model C}. Baseline model and a soft excess, absorbed by a partially covering warm absorber: \texttt{wa$_{\mathrm{G}}$*zxipcf(zpo + bremss+zgauss)}.The free parameters are the same as in model\,A, plus the parameters of the warm absorber. A total of 18\,observations were fitted using this model.\smallskip\newline
{\bf Model D}. Baseline model absorbed by two partially covering warm absorbers: \texttt{wa$_{\mathrm{G}}$*zxipcf*zxipcf(zpo +zgauss)}. The free parameters are the same as in model\,B, with the addition of the parameters of the second ionised absorber ($\xi^2$, $N_{\rm\,H,2}^{\rm\,W}$, $f_{\rm\,W}^2$). Four observations were fitted using this model.\smallskip\newline
{\bf Model E}. Baseline model plus a soft excess and a warm absorber, absorbed by neutral material: \texttt{wa$_{\mathrm{G}}$*zwabs*zxipcf(zpo + bremss+zgauss)}. The free parameters are the same as in model\,C, plus the column density of the neutral absorber ($N_{\rm\,H}^{\rm\,C}$). This model was used for one observation.\smallskip\newline
{\bf Model F}.  Baseline model and a soft excess absorbed by two partially covering warm absorbers: \texttt{wa$_{\mathrm{G}}$*zxipcf*zxipcf(zpo + bremss+zgauss)}. The free parameters are the same as in model\,D, with the addition of the temperature and normalisation of the bremsstrahlung. Three observations were fitted using this model.\smallskip\newline
{\bf Model G}. Baseline model and a soft excess, obscured by a neutral and two partially covering ionised absorbers: \texttt{wa$_{\mathrm{G}}$*zwabs*zxipcf*zxipcf(zpo + bremss+zgauss)}. The free parameters are the same as in model\,D, with the addition of $N_{\rm\,H}^{\rm\,C}$ and of the temperature and normalisation of the bremsstrahlung. This model was used for one observation.\smallskip\newline

For the 14 observations for which it was not possible to constrain the energy of the Fe\,K$\alpha$ line, we fixed the parameter to $E_{\rm\,K\alpha}=6.4\rm\,keV$ (in the rest frame of the AGN). We fixed the width of the Gaussian line to $\sigma=1\rm\,eV$, a value below the energy resolution of EPIC/PN and MOS, in order to only consider the narrow core of the iron K$\alpha$ line. 
We used the values of the Galactic hydrogen column density $N_{\rm\,H}^{\rm\,G}$ obtained by \citet{Dickey:1990uq} mapping the HI emission of the Galaxy (see Table\,\ref{tab:obslog}). 

For all the sources, we tested whether adding a broad component of the iron K$\alpha$ line would significantly improve the fit. This was done using the broad-line profile of \citet{Laor:1991fk} (in XSPEC \texttt{laor2}). Similar to what was done by \citet{Nandra:2007ly}, we fixed the internal (for $r\leq R_{\mathrm{break}}$) emissivity indices to $\beta_1=0$, and the external one (for $r>R_{\mathrm{break}}$) to $\beta_2=3$. The inclination angle was fixed to the value obtained by MIR studies ($i=\theta_{\rm\,i}$). We fixed the energy of the broad line to $E_{\rm\,K\alpha}^{\rm\,broad}=6.4$\,keV (in the reference frame of the AGN) and tried two scenarios: one in which the inner radius of the iron K$\alpha$-emitting region is $r_{\mathrm{in}}=6\,r_{\mathrm{g}}$ (equivalent to the non-rotating Schwarzschild black hole case), where $r_{\mathrm{g}}=GM_{\mathrm{BH}}/c^2$ is the gravitational radius, and the other in which $r_{\mathrm{in}}=1.24\,r_{\mathrm{g}}$ (equivalent to the rotating Kerr black hole scenario). The outer radius of the Fe\,K$\alpha$ emitting region was set in both cases to $r_{\mathrm{out}}=400\,r_{\mathrm{g}}$, while $R_{\mathrm{break}}$ was left as a free parameter. We performed an F-test using the results obtained with and without relativistic Fe\,K$\alpha$ emission, and rejected the presence of a broad line if the probability was $p<95\%$. We found that a broad component of the line is needed for 14 observations and 6 objects (25\% of the total sample). In the following we use only the narrow component of the Fe K$\alpha$ line. For 21 observations additional emission lines (such as O\,VII, Ne\,IX, Fe\,XXV, or Fe\,XXVI) were needed to obtain a good reduced $\chi^{2}$. In Appendix\,\ref{Appendix0} we report the values of the main parameters obtained by our spectral analysis, while all the details of the fits are reported in Appendix\,\ref{Appendix1}. As an example we illustrate in Fig.\,\ref{fig:xrayspec} two typical fits to the X-ray spectra of the sources of our sample.
Several of the PG quasars of our sample have been studied by \cite{Jimenez-Bailon:2005fk} (see also \citealp{Piconcelli:2005ly}), and the values of the EW we obtained are consistent with those reported in their paper.

The flux of the power-law continuum in the 2--10\,keV band ($F_{\mathrm{\,2-10}}$) was obtained using the convolution model \texttt{cflux} in XSPEC. The $k$-corrected continuum luminosities ($L_{\mathrm{\,2-10}}$) were calculated using
\begin{equation}
L_{\mathrm{\,2-10}}=4\pi d_{\rm\,L}^{2}\frac{F_{\mathrm{\,2-10}}}{(1+z)^{2-\Gamma}},
\end{equation}
where $d_{\rm\,L}$ is the luminosity distance. We used standard cosmological parameters ($H_{0}=70\rm\,km\,s^{-1}\,Mpc^{-1}$, $\Omega_{\mathrm{m}}=0.3$, $\Omega_{\Lambda}=0.7$). The iron K$\alpha$ luminosities ($L_{\rm\,K\alpha}$) were calculated in a similar fashion, excluding the $1/(1+z)^{2-\Gamma}$ $k$-correction.

\section{Renormalising the values of EW}\label{Sect:renorm}
\begin{figure}[t!]
\centering
\centering
\includegraphics[width=9cm]{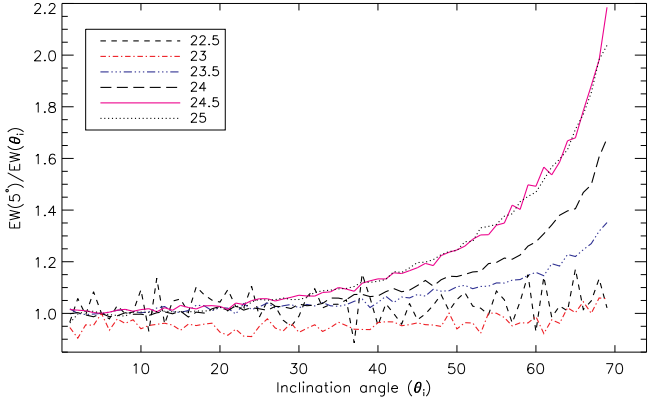}
  \caption{Ratio of the iron K$\alpha$ EW at $\theta_{\mathrm{\,i}}=5^{\circ}$ and the $EW(\theta_{\rm\,i})$ for different values of the equatorial column density of the torus $N_{\rm\,H}^{\rm\,T}$ obtained using the model of \citet{Ikeda:2009nx} for $\theta_{\rm\,OA}=70^{\circ}$ and $\Gamma=1.9$. The scatter in the figure is intrinsic to the Monte Carlo simulations.}
\label{fig:ew_thetaoi}
\end{figure}

There are at least six elements that could affect the value of the Fe\,K$\alpha$ EW and introduce a significant scatter in the correlations with the torus properties obtained by MIR studies: i) \emph{Variability}. The delayed response of the reprocessing material to flux changes of the continuum is expected to have a significant impact on the observed values of EW. ii) The \emph{photon index} of the X-ray emission. Higher values of $\Gamma$ imply a steeper continuum and fewer photons at the energy of the iron K$\alpha$ line, which results in lower values of the EW (e.g., \citealp{Ricci:2013vn}). iii) The \emph{inclination angle} of the observer with respect to the torus. For the geometries considered here, lower values of $\theta_{\rm\,i}$ produce higher values of EW because the observer is able to see more of the reflected flux (e.g., \citealp{Ikeda:2009nx}, see Fig.\,\ref{fig:ew_thetaoi}). iv) The \emph{equatorial column density of the torus} ($N_{\rm\,H}^{\rm\,T}$). EW increases with $N_{\rm\,H}^{\rm\,T}$ up to $\log N_{\rm\,H}^{\rm\,T}\simeq 24$, and above this value is roughly constant (e.g., \citealp{Ghisellini:1994uq}). v) The \emph{half-opening angle of the torus}. The EW of the line decreases for increasing values of $\theta_{\rm\,OA}$  (e.g., \citealp{Ikeda:2009nx}). vi) \emph{Metallicity}. Lower values of the metallicity produce lower values of EW. 

While the impact on the Fe\,K$\alpha$ EW of the first five elements can be reduced, our knowledge of the metallicity of the circumnuclear material of AGN is still poor, so that it is not possible to take this factor into account. In the following, we describe our procedure for renormalising and correcting the values of EW.

\subsection{Variability}
Continuum variability is expected to affect the Fe\,K$\alpha$ EW measured by single observations of AGN. The bulk of the material responsible for the Fe\,K$\alpha$ line is in fact thought to be located several light years from the X-ray source, so that variations in the continuum do not correspond to simultaneous variations in the line emission. This implies that if a source enters a high-flux state, then the flux of the Fe\,K$\alpha$ line relative to that of the continuum (i.e., its EW) is lower than the real value, while it would be higher in a low-flux state. This has been confirmed by the recent work of \citet{Shu:2012fk}, who found an anti-correlation between EW and flux for different observations of the same sources. \citet{Shu:2010zr} show that variability might also play a role in the X-ray Baldwin effect and that the anti-correlation is attenuated when the values of EW are averaged over several observations. To account for this effect we averaged, when possible, the results obtained by different observations of the same source. Since in several cases the same source was observed at an interval of a few days, we averaged all the parameters obtained by observations carried out within one month. Comparing the fluxes of the narrow Fe K$\alpha$ line for the sources for which several observations were available, we found that in almost all cases the fluxes are consistent within the uncertainties. The only exceptions are IC\,4329A and NGC\,4151. The flux of the narrow line in IC\,4329A varies from $1.0^{+0.3}_{-0.4}\times 10^{-12}$ to $5.1^{+0.8}_{-0.6}\times 10^{-13}\rm\,erg\,cm^{-2}\,s^{-1}$ on a time span of 2.5 years, while in NGC\,4151 the line was at its maximum in May 2006 ($2.6^{+0.1}_{-0.1}\times 10^{-12}\rm\,erg\,cm^{-2}\,s^{-1}$) and at its minimum in December 2000 ($1.8^{+0.1}_{-0.1}\times 10^{-12}\rm\,erg\,cm^{-2}\,s^{-1}$).
The low variability of the flux of the narrow Fe K$\alpha$ line agrees with the idea that the cold material where the X-ray radiation is reprocessed is located far away from the X-ray source.

\subsection{Removing the dependence on $\Gamma$, $\theta_{\rm\,i}$, and $N_{\rm\,H}^{\rm{\,T}}$}\label{sect:corrections}
Thanks to the advent of physical torus models such as those developed by \citeauthor{Murphy:2009uq} (\citeyear{Murphy:2009uq}; \texttt{MYTorus}\footnote{http://www.mytorus.com/}), \citet{Ikeda:2009nx}, and \citet{Brightman:2011oq}, it is now possible to reproduce the reflection features produced in the dusty torus surrounding the X-ray source. These models can also be used to remove the dependence of EW on $\Gamma$, $\theta_{\rm\,i}$, and $N_{\rm\,H}^{\mathrm{\,T}}$, renormalising EW to the same set of values for each source. Removing the degeneracy on EW introduced by these parameters is important to assess the real dependence of EW on the covering factor of the torus (and thus on $\theta_{\rm\,OA}$). This can be done by using the average values of the photon index obtained by the X-ray spectral fitting ($\Gamma^{\,\mathrm{obs}}$), and the values of the inclination angle ($\theta_{\rm\,i}^{\mathrm{\,obs}}$) and half-opening angle of the torus ($\theta_{\rm\,OA}^{\mathrm{\,obs}}$) obtained by the MIR analysis. The values of the equatorial column density of the torus ($N_{\rm\,H}^{\mathrm{\,T\,,obs}}$) can be extrapolated from the results of the MIR spectral fitting. From the relation between the extinction in the $V$ band and the optical depth ($A_{\mathrm{V}}=1.086\,\tau_{\mathrm{V}}$), and from the Galactic relation between extinction and column density found by {\it ROSAT} ($N_{\rm\,H}=A_{\mathrm{V}}\cdot 1.79\times 10^{21}\rm\,cm^{-2}$, \citealp{Predehl:1995ly}), for the torus one obtains:
\begin{equation}\label{Eq:nhtcalc}
N_{\rm\,H}^{\mathrm{\,T,\,obs}}=1.086\,N_{0}\cdot\tau_{\mathrm{V}}\cdot1.79\times 10^{21}\rm\,cm^{-2}.
\end{equation}

Equation\,\ref{Eq:nhtcalc} assumes a Galactic $A_{\mathrm{V}}/N_{\rm\,H}$ ratio. This might, however, represent a crude approximation of the real value. \cite{Maiolino:2001oq,Maiolino:2001nx} have shown that the $E(B-V)/N_{\rm\,H}$ ratio in AGN (for $\log L_{2-10}\geq 42$) ranges from $\sim 1\%$ to $\sim 40\%$ of the Galactic value. As discussed by \cite{Maiolino:2001nx}, this is likely to imply that the $A_{\mathrm{V}}/N_{\rm\,H}$ ratio is also significantly lower in AGN than in our Galaxy. 
To take this effect into account, we used the average value of the $E(B-V)/N_{\rm\,H}$ ratio found by \cite{Maiolino:2001oq} (for $\log L_{2-10}\geq 42$) to recalculate the equatorial column density of the torus ($N_{\rm\,H,\,Av.}^{\mathrm{\,T,\,obs}}$). To convert the values of $E(B-V)/N_{\rm\,H}$ into $A_{\mathrm{V}}/N_{\rm\,H}$, we adopted the Galactic ratio ($A_{\mathrm{V}}/E(B-V)=3.1$). The value of $N_{\rm\,H,\,Av.}^{\mathrm{\,T,\,obs}}$ is then calculated similarly to what was done in Eq.\,\ref{Eq:nhtcalc}: 

\begin{equation}\label{Eq:nhtcalcav}
N_{\rm\,H,\,Av.}^{\mathrm{\,T,\,obs}}=1.086\,N_{0}\cdot\tau_{\mathrm{V}}\cdot1.1\times 10^{22}\rm\,cm^{-2}.
\end{equation}
In the following we consider both values of $N_{\rm\,H}^{\mathrm{\,T}}$ obtained using Eqs.\,\ref{Eq:nhtcalc} and \ref{Eq:nhtcalcav}. The values of $f_{\rm\,obs}$ and $N_{\rm\,H}^{\mathrm{\,T,\,obs}}$ are reported in Table\,\ref{tab:fitresults}.

Simulations of X-ray absorption and reflection from a clumpy torus have not been carried out yet, and the models listed above consider a smooth dust distribution, different from that of the model of \citet{Nenkova:2008uq,Nenkova:2008kx}. However, assuming that most of the Fe K$\alpha$ line is produced in the outer skin of the torus, the differences between the two geometries are likely to be small for type-I AGN. To correct the values of Fe\,K$\alpha$ EW, we used the model developed by \citet{Ikeda:2009nx}. This model considers a spherical-toroidal geometry for the reprocessing material and has the advantage, with respect to \texttt{MYTorus}, of having the half-opening angle of the torus $\theta_{\rm\,OA}$ as a free parameter. The model of \citet{Brightman:2011oq} assumes a similar geometry, but considers a less realistic line-of-sight column density, which is constant for all values of $\theta_{\rm\,i}$. The other free parameters of the model of \citet{Ikeda:2009nx} are $\Gamma$, $N_{\rm\,H}^{\mathrm{\,T}}$, and $\theta_{\rm\,i}$. Using the model of \citet{Ikeda:2009nx}, we simulated, for each source, a spectrum with the parameters fixed to the values obtained by the observations, and one with a set of parameters arbitrarily chosen ($\Gamma=1.9,\theta_{\rm\,i}=5^{\circ},N_{\rm\,H}^{\mathrm{\,T}}=10^{24}\rm\,cm^{-2}$) and with $\theta_{\rm\,OA}$ fixed to the value obtained by the fit to the MIR spectra ($\theta_{\rm\,OA}^{\mathrm{\,obs}}$). We obtained the value of the Fe K$\alpha$ EW of the simulated spectrum ($EW_{\rm\,mod}$) following what was done in \citet{Ricci:2013fk}. For each source we calculated the corrections $K _{\mathrm{I}}$ with

\begin{equation}\label{Eq:corr_ik}
K _{\mathrm{I}}=\frac{EW_{\rm\,mod}(\Gamma=1.9,\theta_{\rm\,i}=5^{\circ},N_{\rm\,H}^{\mathrm{\,T}}=10^{24}\rm\,cm^{-2},\theta_{\rm\,OA}^{\mathrm{\,obs}})}{EW_{\rm\,mod}(\Gamma^{\,\mathrm{obs}},\theta_{\rm\,i}^{\mathrm{\,obs}},N_{\rm\,H}^{\mathrm{\,T,\,obs}},\theta_{\rm\,OA}^{\mathrm{\,obs}})}.
\end{equation}

The renormalised equivalent widths ($EW_{\mathrm{corr}}^{\rm\,I}$) can be easily calculated from the corrections and the observed values of the equivalent width ($EW_{\rm\,obs}$):
\begin{equation}\label{Eq:corr}
EW_{\mathrm{corr}}^{\rm\,I}=K_{\mathrm{I}}\times EW_{\rm\,obs}.
\end{equation}
As an example of the corrections used, in Fig.\,\ref{fig:ew_thetaoi} we report the trend of EW($5^{\circ}$)/EW($\theta_{\rm\,i}$) for different values of $N_{\rm\,H}^{\mathrm{\,T}}$.

\subsection{Removing the dependence on $\Gamma$, $\theta_{\rm\,i}$, and $\theta_{\rm\,OA}$}\label{sect:corrections2}
To study the intrinsic relation between the Fe\,K$\alpha$ EW and the equatorial column density of the torus, one can use a procedure similar to the one adopted in Sect.\,\ref{sect:corrections}. Since we are now interested in $N_{\rm\,H}^{\mathrm{\,T}}$, we fixed the half-opening angle to an arbitrary value ($\theta_{\rm\,OA}=30^{\circ}$) for the reference value of EW and set $N_{\rm\,H}^{\mathrm{\,T}}$ to its observed value. The corrections $K _{\mathrm{II}}$ become
\begin{equation}\label{Eq:corr_ik2}
K _{\mathrm{II}}=\frac{EW(\Gamma=1.9,\theta_{\rm\,i}=5^{\circ},N_{\rm\,H}^{\mathrm{\,T}}=N_{\rm\,H}^{\mathrm{\,T\,,obs}},\theta_{\rm\,OA}=30^{\circ})}{EW(\Gamma^{\,\mathrm{obs}},\theta_{\rm\,i}^{\mathrm{\,obs}},N_{\rm\,H}^{\mathrm{\,T,\,obs}},\theta_{\rm\,OA}^{\mathrm{\,obs}})},
\end{equation}
while the renormalised equivalent width ($EW_{\mathrm{corr}}^{\,II}$) can be obtained by
\begin{equation}\label{Eq:corr2}
EW_{\mathrm{corr}}^{\rm\,II}=K_{\mathrm{II}}\times EW_{\rm\,obs}.
\end{equation}

\section{The relation between the Fe\,K$\alpha$ EW and the properties of the molecular torus}\label{Sect:EWvsf2}
\subsection{Covering factor}\label{section:ewvscf}
As discussed in \citet{Mor:2009fk}, the real covering factor of the torus ($f_{\rm\,obs}$) should be calculated by taking the number of clouds, the half-opening angle of the torus, and the inclination angle of the AGN into account. In the clumpy torus model the probability that the radiation from the central source escapes the torus at a given angle $\beta$ without interacting with the obscuring material is
\begin{equation}\label{eq:prob_esc}
P_{\mathrm{esc}}(\beta)=e^{-N_{\mathrm{0}}\mathrm{exp}\left(-\frac{\beta ^2}{\sigma _{\rm\,tor} ^2}\right)},
\end{equation}
where $\beta=\pi/2-\theta_{\rm\,i}$. The geometrical covering factor of the molecular torus is given by integrating $P_{\mathrm{esc}}$ over all angles
\begin{equation}\label{eq:frac_obsc}
f_{\rm\,obs}=1-\int_{0}^{\pi/2}P_{\mathrm{esc}}(\beta)\cos(\beta)\mathrm{d} \beta.
\end{equation}
\begin{figure}[t!]
\centering
\centering
\includegraphics[width=9cm]{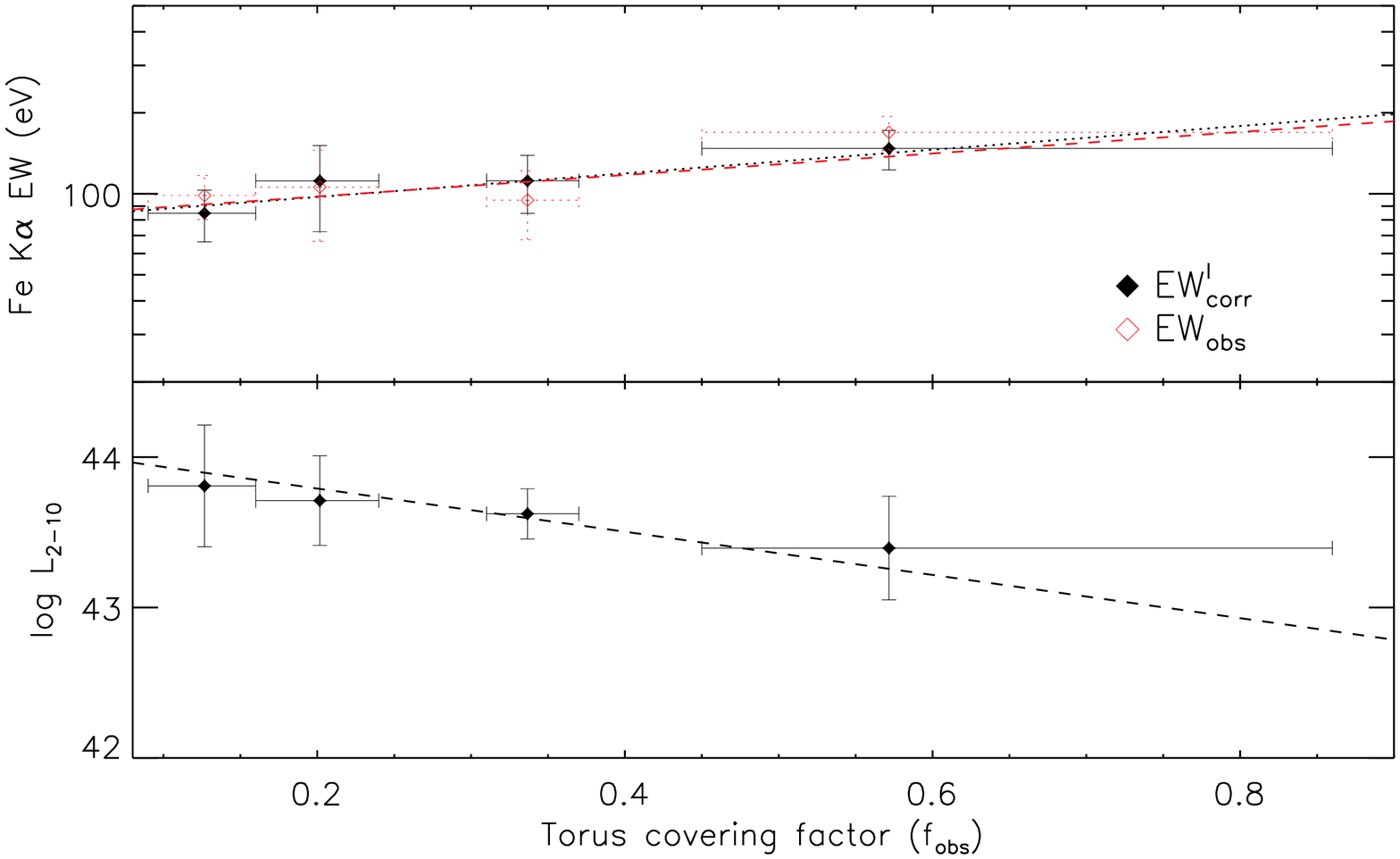}
  \caption{{\it Top panel}: values of the Fe\,K$\alpha$ EW versus the geometrical covering factor of the torus ($f_{\rm\,obs}$) obtained by fitting Mid-IR spectra with the clumpy torus model.  The black filled diamonds are the corrected values (obtained using the model of \citealp{Ikeda:2009nx}, see Eqs.\,\ref{Eq:corr_ik} and \ref{Eq:corr}), while the empty red ones are the uncorrected values multiplied by an arbitrary constant factor for comparison. The data were rebinned to have six values per bin, and the uncertainties on the Fe K$\alpha$ EW were calculated using the standard error of the mean. The dotted line represents the best fit to the non-binned data obtained by applying Eq.\,\ref{eq:EWcorrvsf2}, and it has a slope of $\overline{B}=0.44\pm0.21$. The red dashed line represents the expected $EW-f_{\rm\,obs}$ trend for the set of parameters chosen for the renormalisation, calculated using the model of \citet{Ikeda:2009nx}. The intercept of the expected trend was obtained by fitting the data with the slope fixed to the expected value ($B_{\rm\,exp}\simeq0.4$). {\it Bottom panel}: 2--10\,keV luminosities versus covering factor of the torus for our sample. The dashed line represents the best fit to the data (Eq.\,\ref{eq:logLcvsFobs}). }
\label{fig:ew_f2}

\end{figure}
If the Fe\,K$\alpha$ line is produced in the torus and the results obtained by applying the clumpy torus model to the MIR spectra of AGN are correct, then a positive correlation between the EW of the line and the real covering factor of the torus would be expected. To study the relation between Fe\,K$\alpha$ EW and $f_{\rm\,obs}$ for our sample, which includes several upper limits, we followed the approach of \citet{Guainazzi:2006fk} and \citet{Bianchi:2007vn}, which is an extension of the regression method for left-censoring data described by \citet{Schmitt:1985kx} and \citet{Isobe:1986uq}. We performed 10\,000 Monte-Carlo simulations for each value of the Fe K$\alpha$ EW, taking the two following requirements into account: i) the values of EW of the detections were substituted with a random Gaussian distribution, whose mean is given by the value obtained by the fit, and the standard deviation by its error; ii) the upper limits U were substituted with a random uniform distribution in the interval [0,U]. To reduce the degeneracy introduced by different values of $\Gamma$, $\theta_{\rm\,i}$ and $N_{\rm\,H}^{\rm\,T}$, we used the values of the Fe\,K$\alpha$ EW corrected as described in Sect.\,\ref{sect:corrections}. For each Monte-Carlo run we fitted the values with a log-linear relationship of the type
\begin{equation}\label{eq:EWcorrvsf2}
\log EW_{\mathrm{corr}}^{\rm\,I}=A+B\cdot f_{\rm\,obs},
\end{equation}
using the ordinary least squares (OLS[Y|X]) method. We used the average value of the simulations ($\overline{B}$) as a slope, and as uncertainty their standard deviation. To quantify the significance of the correlation, for each simulation we calculated the Spearman's rank coefficient ($\rho$) and the null hypothesis probability of the correlation ($P_{\mathrm{\,n}}$), and used the values averaged over all the simulations. Applying Eq.\,\ref{eq:EWcorrvsf2}, we obtained a slope of $\overline{B}=0.44\pm0.21$. With the model of \citet{Ikeda:2009nx}, it is possible to deduce the expected $EW-f_{\rm\,obs}$ trend for the set of parameters we used to renormalise the values of EW. The correct formulation of $f_{\rm\,obs}$ is given by Eq.\,\ref{eq:frac_obsc}, but for $\log N_{\rm\,H}^{\rm\,T}=24$ at 6.4\,keV, the escaping probability is $P_{\rm\,esc}\sim 0.08$ for $\beta < \sigma_{\rm\,tor}$, so that we can approximate the relation to $f_{\rm\,obs}\simeq \cos \theta_{\rm\,OA}$. We found that for $\Gamma=1.9$, $N_{\rm\,H}^{\rm\,T}=10^{24}\rm\,cm^{\rm\,-2}$, and $\theta_{\rm\,i}=5^{\circ}$, the expected slope is $B_{\rm\,exp}\simeq 0.4$, consistent with the result of the fit. The scatter plot of $EW_{\mathrm{corr}}^{\rm\,I}$ and $EW_{\rm\,obs}$ versus $f_{\rm\,obs}$ is illustrated in the top panel of Fig.\,\ref{fig:ew_f2}. For graphical clarity the data were rebinned to have six values per bin. Performing the statistical tests described above, we found that however the correlation is statistically not significant, with a null hypothesis probability of $P_{\mathrm{\,n}}=35\%$ and a Spearman's rank coefficient of $\rho=0.22$. Correcting the values of EW using the equatorial column density of the torus obtained by assuming the average $E(B-V)/N_{\rm\,H}$ ratio of \cite{Maiolino:2001nx} does not alter significantly the results ($\overline{B}=0.46\pm0.23$, $\rho=0.23$, $P_{\mathrm{\,n}}=37\%$).

We took random Gaussian errors on $f_{\rm\,obs}$ into account, as done for EW, using the errors reported in \citet{Alonso-Herrero:2011zr} and \citet{Ramos-Almeida:2011ly} and considering uncertainties of $30\%$ for the sources of \citet{Mor:2009fk}. This does not increase the significance of the correlation, giving a null hypothesis probability of $P_{\mathrm{\,n}}=38\%$. We verified whether the fact that the MIR fitting procedures of \citet{Mor:2009fk}, \citet{Alonso-Herrero:2011zr}, and \citet{Ramos-Almeida:2011ly} differ might alter the results. We fitted the data taking only the 19 sources from \citet{Mor:2009fk} into account, and found that the correlation is still not significant ($P_{\mathrm{\,n}}=40\%$). Consistent results ($P_{\rm\,n}=34\%$, $\rho=0.24$) were obtained not considering the observations affected by pile-up in the fit.

In the bottom panel of Fig.\,\ref{fig:ew_f2} we show the scatter plot of the 2--10\,keV luminosity versus the covering factor of the torus. The two parameters are not significantly correlated ($P_{\mathrm{\,n}}=33\%$), and by fitting the data we obtained 
\begin{equation}\label{eq:logLcvsFobs}
\log L_{\,2-10}\propto (-1.44\pm0.78)f_{\rm\,obs}.
\end{equation}

\subsection{Equatorial column density of the torus}\label{section:ewvsnht}

\begin{figure}[t!]
\centering
\centering
\includegraphics[width=9cm]{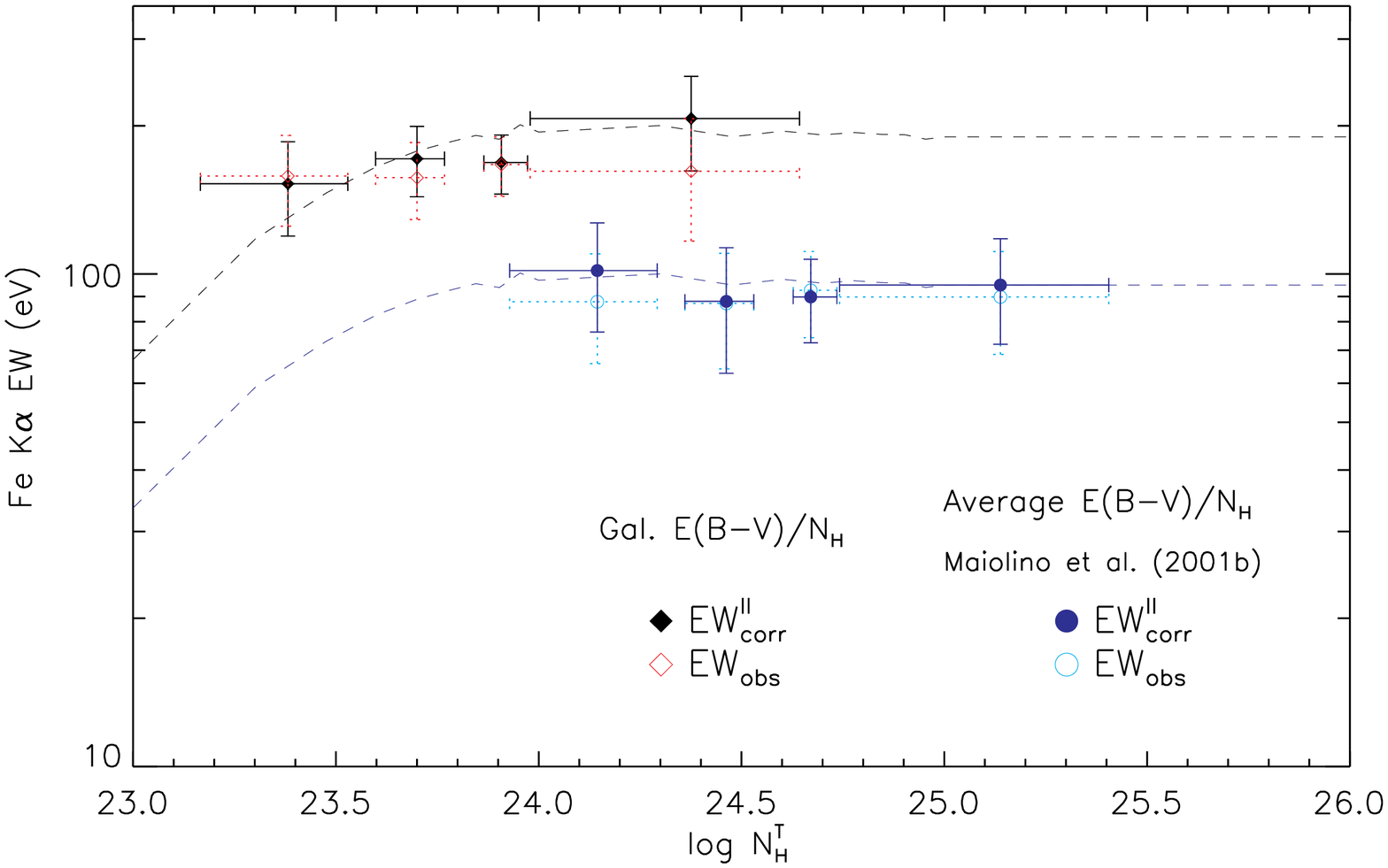}
  \caption{Values of the Fe\,K$\alpha$ EW versus the equatorial column density of the torus ($N_{\rm\,H}^{\rm\,T}$) obtained by fitting the mid-IR spectra. Diamonds and circles represent values of $N_{\rm\,H}^{\rm\,T}$ obtained using the Galactic $E(B-V)/N_{\rm\,H}$ ratio and the average value of \cite{Maiolino:2001nx}, respectively. The black (blue) filled diamonds (circles) are the corrected values (obtained using the model of \citealp{Ikeda:2009nx}, see Eqs.\,\ref{Eq:corr_ik2} and \ref{Eq:corr2}), while the empty red (cyan) diamonds (circles) are the uncorrected values. The values of EW were multiplied by an arbitrary constant factor for comparison. The data were rebinned to have six values per bin, and the uncertainties on the Fe K$\alpha$ EW are calculated using the standard error of the mean. The dashed lines represent the expected $EW-N_{\rm\,H}^{\rm\,T}$ trend for the set of parameters chosen for the renormalisation, calculated using the model of \citet{Ikeda:2009nx}. The expected trends were normalised to be compatible with the corrected data.}
\label{fig:ew_NHT}

\end{figure}

With the values of the Fe\,K$\alpha$ EW corrected to remove the dependence on $\Gamma$, $\theta_{\rm\,i}$, and $\theta_{\rm\,OA}$ (Sect.\,\ref{sect:corrections2}), we searched for a correlation with the equatorial column density of the torus. Monte Carlo simulations (e.g., \citealp{Ikeda:2009nx}, \citealp{Murphy:2009uq}) have shown that this parameter is expected to play an important role on the iron K$\alpha$ line EW. Following the same procedure as discussed in Sect.\,\ref{section:ewvscf}, we found that the correlation is statistically not significant ($\rho=0.25$, $P_{\rm\,n}=29\%$). Fitting the data with a log-linear relation of the type  
\begin{equation}\label{eq:EWvsNHT}
\log EW_{\mathrm{corr}}^{\rm\,II}= \alpha+ \beta \cdot \log N_{\rm\,H}^{\rm\,T},
\end{equation}
we obtained $\overline{\beta}=0.16\pm0.08$.  However, due to self-absorption for large values of the column density of the torus, the Fe\,K$\alpha$ EW is expected to saturate for $\log N_{\rm\,H}^{\rm\,T} \gtrsim 24$ (e.g., \citealp{Ghisellini:1994uq}), so that a linear increment is expected only up to this value. Considering only the data for $\log N_{\rm\,H}^{\rm\,T} \leq 24$ resulted in a slope ($\overline{\beta}=0.32\pm0.23$) which is consistent to the expected value ($\beta_{\rm\,exp}=0.53$) for the same range of $N_{\rm\,H}^{\rm\,T}$. The trend is, however, statistically non-significant ($\rho=0.30$, $P_{\rm\,n}=31\%$). In Fig.\,\ref{fig:ew_NHT} we show the scatter plot of $EW_{\mathrm{corr}}^{\rm\,II}$ and $EW_{\mathrm{obs}}$ versus $N_{\rm\,H}^{\rm\,T}$, together with the expected trend calculated using the model of \citet{Ikeda:2009nx} for the set of parameters used for the re-normalisation. Both the values of $EW_{\mathrm{corr}}^{\rm\,II}$ and $EW_{\mathrm{obs}}$ agree with the predicted trend, as would be expected if the line was produced in the molecular torus, although their large associated uncertainties do not allow us to draw a firm conclusion.

\section{Summary and discussion}\label{Sect:summary}

Reflection of the power-law continuum from circumnuclear material in AGN is mainly observed through the narrow Fe\,K$\alpha$ line and a reflection hump peaking at $\sim 30$\,keV. The fraction of continuum X-ray flux reflected (hence the Fe\,K$\alpha$ EW) is likely to depend strongly on the covering factor of the torus ($f_{\rm\,obs}$). This implies that $f_{\rm\,obs}$, and its evolution with the physical properties of the AGN, is fundamental for a correct understanding of the cosmic X-ray background (CXB, e.g., \citealp{Gilli:2007qf}). The maximum emission of the CXB is in fact observed at $\sim30$\,keV (e.g., \citealp{Marshall:1980kx}), and a large fraction of CT AGN has often been invoked to correctly reproduce its shape (e.g., \citealp{Gilli:2007qf}). Because it is observed through a large amount of obscuring material, most of the continuum in these objects is absorbed, which enhances the apparent reflected-to-incident flux ratio, and it makes their observed spectra peak at $\sim 30$\,keV. However, the fraction of Compton-thick sources needed to explain the peak is strongly linked to the fraction of reflected continuum \citep{Gandhi:2007fk,Treister:2009uq}, and thus to $f_{\rm\,obs}$. High values of $f_{\rm\,obs}$ have been invoked to explain the characteristics of buried AGN \citep{Ueda:2007fk,Eguchi:2009kx,Eguchi:2011uq}, which are type-II objects that have a strong reflection component and a low fraction of scattered continuum. A large covering factor of the torus might explain the strong reflection observed in the hard X-ray spectrum of mildly obscured ($23 \leq \log N_{\rm\,H} < 24$) AGN found by stacking {\it INTEGRAL} IBIS/ISGRI data \citep{Ricci:2011vn} and recently confirmed by \citet{Vasudevan:2013ys} using {\it Swift}/BAT. The covering factor of the torus is believed to decrease with luminosity. The original idea  of such a relation was put forward by \citet{Lawrence:1982bh} to explain the decrease in the fraction of obscured sources with the luminosity. In the past decade, this trend has been confirmed by several studies carried out at different wavelengths (e.g., \citealp{Ueda:2003qf}, \citealp{Beckmann:2009ys}), and it has been shown that it would also be able to straightforwardly explain the X-ray Baldwin effect \citep{Ricci:2013fk}. Thus a correct understanding of the relation between reflected X-ray radiation and the covering factor of the torus is of the utmost importance for a complete understanding of the X-ray spectral evolution of AGN.

In this work we have studied the relation between the Fe\,K$\alpha$ EW and important physical properties of the molecular torus, such as its covering factor and equatorial column density ($N_{\rm\,H}^{\rm\,T}$), for a sample of 24 AGN. This was done by combining {\it XMM-Newton}/EPIC observations in X-rays with the results obtained by recent MIR spectral studies carried out using the clumpy torus models of \citet{Nenkova:2008uq,Nenkova:2008kx}. The physical torus model of \citet{Ikeda:2009nx} was used to correct the values of the Fe\,K$\alpha$ EW, in order to remove the degeneracy introduced by different values of $\Gamma$, $\theta_{\rm\,i}$, and $N_{\rm\,H}^{\rm\,T}$. We found that, although the correlation between the Fe\,K$\alpha$ EW and the covering factor of the torus is statistically non-significant, the slope obtained ($\overline{B}=0.44\pm0.21$) is in very good agreement with the expected value ($B_{\rm\,exp}\simeq 0.4$, see Fig.\,\ref{fig:ew_f2}). A similar result is obtained when studying the relation between Fe\,K$\alpha$ EW and $N_{\rm\,H}^{\rm\,T}$ for $\log N_{\rm\,H}^{\rm\,T} \leq 24$: the slope obtained ($\overline{\beta}=0.32\pm0.23$) is consistent with the predicted value ($\overline{\beta}_{\rm\,exp}=0.53$, see Fig.\,\ref{fig:ew_NHT}), although the correlation is statistically non-significant.

The fact that the correlation between EW and $f_{\rm\,obs}$ is statistically not significant is probably related to the large errors of Fe\,K$\alpha$ EW, and to the large number of PG quasars in the sample, which skews the luminosity distribution towards high values. We also cannot exclude the effect of systematic errors introduced by the technique used to fit the MIR spectra. As argued by \citet{Mor:2009fk}, there are two main uncertainties associated to their treatment: variability between the non-simultaneous optical and MIR observations and their choice of the bolometric corrections. These uncertainties could introduce scatter into the torus parameters obtained, so that larger samples, with a more uniform luminosity distribution, are probably needed to find a clear trend between Fe\,K$\alpha$ EW, $f_{\rm\,obs}$, and $N_{\rm\,H}^{\rm\,T}$. 
Another possible source of uncertainty could be introduced by the fact that the model of \citet{Nenkova:2008uq,Nenkova:2008kx} assumes the inner radius of the torus given by
\begin{equation}\label{eq:inner_radius_torus}
R_{\rm\,d}=0.4 \left(\frac{L_{\rm\,bol}}{10^{\,45}\rm\,erg\,s^{-1}}\right)^{1/2}\left(\frac{1500\,K}{T_{\rm\,sub}}\right)\rm\,pc,
\end{equation}
where $L_{\rm\,bol}$ is the bolometric luminosity of the AGN and $T_{\rm\,sub}$ is the dust sublimation temperature. NIR reverberation studies of the torus have shown that Eq.\,\ref{eq:inner_radius_torus} overestimates the value of $R_{\rm\,d}$, which is found to be systematically smaller by a factor of three \citep{Kishimoto:2007zr}. \citet{Kawaguchi:2010fk} show that the discrepancy is probably related to the fact that the accretion disk emits anisotropically, so that the effective inner radius is smaller than predicted by Eq.\,\ref{eq:inner_radius_torus}. Furthermore, it has been suggested that the observed NIR excess may be due to the fact that silicate and graphite grains sublimate at different temperatures and that large grains are cooling more efficiently, leading to a sublimation zone rather than a sublimation radius (e.g., \citealp{Kishimoto:2007zr}, \citealp{Mor:2012ys}). Although the hot graphite-only zone is a plausible source of an additional NIR emission, this component still has to be consistently included in the radiative transfer modelling. \cite{Schartmann:2009ve} models do account for the latter effect by separating the grains of different sizes, but no NIR bump is seen. Other suggestions for the source of the NIR excess include an additional component of low-density interclump dust \citep{Stalevski:2012fk,Stalevski:2013ly} or assume that NIR and MIR emission are coming from two spatially very distinct regions \citep{Honig:2013bh}. Moreover, modelling of the dusty tori IR emission comes with caveats of its own \citep{Hoenig:2013qf}, and model parameters are often degenerate, sometimes resulting in similarly good fits for different combinations of parameters. This inevitably introduces additional uncertainties in any analysis that relies on the properties of the torus obtained from fitting their IR SEDs.

An additional source of scatter might be related to the fact that Fe\,K$\alpha$ emission originating in the torus is subject to a significant contamination from other regions of the AGN, such as the BLR or the outer part of the accretion disk. This degeneracy will be broken in a few years with the advent of {\it ASTRO-H} \citep{Takahashi:2010uq}. Thanks to the unprecedented energy resolution in the Fe\,K$\alpha$ energy band of its X-ray calorimeter (SXS, 5\,eV FWHM at 6\,keV), {\it ASTRO-H} will be able to disentangle the emission produced in the torus from that arising from different regions of the AGN.

\begin{acknowledgements}
We thank the anonymous referee for his/her comments that helped improve the paper. We thank Chin Shin Chang and Almudena Alonso Herrero for their comments on the manuscript. CR is a Fellow of the Japan Society for the Promotion of Science (JSPS). This work was partly supported by the Grant-in-Aid for Scientific Research 23540265 (YU) from the Ministry of Education, Culture, Sports, Science and Technology of Japan (MEXT). 
PG acknowledges support from STFC grant reference ST/J00369711. MS acknowledges support of the Ministry of Education, Science and Technological Development of the Republic of Serbia through the projects Astrophysical Spectroscopy of Extragalactic Objects (176001) and Gravitation and the Large Scale Structure of the Universe (176003), and by FONDECYT through grant No. 3140518. This research has made use of the NASA/IPAC Extragalactic Database (NED) which is operated by the Jet Propulsion Laboratory, of data obtained from the High Energy Astrophysics Science Archive Research Center (HEASARC), provided by NASA's Goddard Space Flight Center, and of the SIMBAD Astronomical Database, which is operated by the Centre de Donn\'ees astronomiques de Strasbourg. Based on observations obtained with {\it XMM-Newton}, an ESA science mission with instruments and contributions directly funded by ESA Member States and NASA.

 \end{acknowledgements}

\longtab{2}{
\begin{landscape}
\begin{center}
\LTcapwidth=1.13\textwidth
\footnotesize
\begin{longtable}{lcccccccccccccccc}

\caption{Results of the spectral fits of the {\it XMM-Newton} observations of our sample. (1) Model used for the fit (for details see Table\,\ref{tab:fitresults2} and Sect.\,\ref{Sect:spectral_analysis}), (2) photon index of the power-law continuum, (3) flux and (4) luminosity of the 2--10\,keV power-law continuum, (5) energy, (6) equivalent width, (7) flux, and (8) luminosity of the narrow component of the iron K$\alpha$ line, (9) probability of the presence of a broad relativistic component of the iron K$\alpha$ line and (10) its equivalent width. The table also lists the characteristics of the molecular torus obtained by fitting mid-IR spectra of AGN: (11) covering factor and (12) equatorial column density of the torus (calculated using Eq.\,\ref{Eq:nhtcalc}, the values are $\sim 6$ times higher if calculated with Eq.\,\ref{Eq:nhtcalcav}), (13) bibliographical reference for the value.}\label{tab:fitresults}\\
\hline \hline \\


\multicolumn{1}{l}{ Source} & 
\multicolumn{4}{c}{ 2--10\,keV continuum} &  \phantom{} &
\multicolumn{4}{c}{Narrow Fe\,K$\alpha$ line} &  \phantom{} &
\multicolumn{2}{c}{Broad Fe\,K$\alpha$ line}  &  \phantom{} &
\multicolumn{3}{c}{Molecular torus} \\ 

\multicolumn{1}{l}{  } & 
\multicolumn{4}{l}{ } &  \phantom{} &
\multicolumn{4}{l}{} &  \phantom{} &
\multicolumn{2}{l}{}  &  \phantom{} &
\multicolumn{3}{l}{}  \\ \cline{2-5} \cline{7-10} \cline{12-13} \cline{15-17}

\noalign{\smallskip}
\noalign{\smallskip}

   \multicolumn{1}{l}{ } & 
   \multicolumn{1}{c}{(1) } & 
   \multicolumn{1}{c}{(2) } & 
   \multicolumn{1}{c}{(3) } & 
   \multicolumn{1}{c}{ (4)} &  \phantom{} &
   \multicolumn{1}{c}{(5) } & 
   \multicolumn{1}{c}{(6) } & 
      \multicolumn{1}{c}{ (7)} & 
   \multicolumn{1}{c}{(8) } &  \phantom{} &
   \multicolumn{1}{c}{(9) } & 
   \multicolumn{1}{c}{(10) } &  \phantom{} &
   \multicolumn{1}{c}{(11) } &
   \multicolumn{1}{c}{(12) } &
   \multicolumn{1}{c}{(13) } 
    \\
\noalign{\smallskip}

   \multicolumn{1}{l}{ } & 
   \multicolumn{1}{c}{Model } & 
   \multicolumn{1}{c}{ $\Gamma$} & 
   \multicolumn{1}{c}{$\log f_{\mathrm{\,2-10}}$ } & 
   \multicolumn{1}{c}{$\log L_{\mathrm{\,2-10}}$ } &  \phantom{} &
   \multicolumn{1}{c}{ $E_{\rm\,K\alpha} $ } & 
   \multicolumn{1}{c}{EW } & 
   \multicolumn{1}{c}{ $\log f_{\mathrm K\alpha}$} & 
   \multicolumn{1}{c}{ $\log L_{\rm\,K\alpha}$} &  \phantom{} &
   \multicolumn{1}{c}{p } & 
   \multicolumn{1}{c}{EW } &  \phantom{} &
   \multicolumn{1}{c}{ $f_{\rm\,obs}$ } &
   \multicolumn{1}{c}{ $\log N_{\rm\,H}^{\mathrm{\,T,\,obs}}$ } &
   \multicolumn{1}{c}{ Ref. }    \\ 

\noalign{\smallskip}

   \multicolumn{1}{l}{ } & 
   \multicolumn{1}{c}{  } & 
   \multicolumn{1}{c}{ } & 
   \multicolumn{1}{c}{[{\tiny $\rm\,erg\,cm^{-2}\,s^{-1}$}]  } & 
   \multicolumn{1}{c}{ [{\tiny $\rm\,erg\,s^{-1}$}]} &  \phantom{} &
   \multicolumn{1}{c}{[{\tiny keV}] } & 
   \multicolumn{1}{c}{[{\tiny eV}]} & 
      \multicolumn{1}{c}{ [{\tiny $\rm\,erg\,cm^{-2}\,s^{-1}$}] } & 
   \multicolumn{1}{c}{[{\tiny $\rm\,erg\,s^{-1}$}] } &  \phantom{} &
   \multicolumn{1}{c}{  } & 
   \multicolumn{1}{c}{ [{\tiny eV}] }  &  \phantom{} &
   \multicolumn{1}{c}{  } & 
   \multicolumn{1}{c}{ [{\tiny $\rm\,cm^{-2}$}]  } & 
   \multicolumn{1}{c}{ }       \\
   
\noalign{\smallskip}

\noalign{\smallskip}

\endfirsthead

\multicolumn{17}{l}%
{\small \hspace{-0.1in}{{\textbf{\tablename\ \thetable{}.} Results of the spectral fits of the {\it XMM-Newton} observations of our sample. -- \textit{continued} }}} \\
\noalign{\smallskip}
\hline \hline \\[0.05ex]


\multicolumn{1}{l}{ Source} & 
\multicolumn{4}{c}{ 2--10\,keV continuum} &  \phantom{} &
\multicolumn{4}{c}{Narrow Fe\,K$\alpha$ line} &  \phantom{} &
\multicolumn{2}{c}{Broad Fe\,K$\alpha$ line}  &  \phantom{} &
\multicolumn{3}{c}{Molecular torus} \\ 

\multicolumn{1}{l}{  } & 
\multicolumn{4}{l}{ } &  \phantom{} &
\multicolumn{4}{l}{} &  \phantom{} &
\multicolumn{2}{l}{}  &  \phantom{} &
\multicolumn{3}{l}{} 
 \\ \cline{2-5} \cline{7-10} \cline{12-13} \cline{15-17}

\noalign{\smallskip}

\noalign{\smallskip}

   \multicolumn{1}{l}{ } & 
   \multicolumn{1}{c}{(1) } & 
   \multicolumn{1}{c}{(2) } & 
   \multicolumn{1}{c}{(3) } & 
   \multicolumn{1}{c}{ (4)} &  \phantom{} &
   \multicolumn{1}{c}{(5) } & 
   \multicolumn{1}{c}{(6) } & 
      \multicolumn{1}{c}{ (7)} & 
   \multicolumn{1}{c}{(8) } &  \phantom{} &
   \multicolumn{1}{c}{(9) } & 
   \multicolumn{1}{c}{(10) } &  \phantom{} &
   \multicolumn{1}{c}{(11) } &
   \multicolumn{1}{c}{(12) } &
   \multicolumn{1}{c}{(13) }    \\
\noalign{\smallskip}

   \multicolumn{1}{l}{ } & 
   \multicolumn{1}{c}{Model } & 
   \multicolumn{1}{c}{ $\Gamma$} & 
   \multicolumn{1}{c}{$\log f_{\mathrm{\,2-10}}$ } & 
   \multicolumn{1}{c}{$\log L_{\mathrm{\,2-10}}$ } &  \phantom{} &
   \multicolumn{1}{c}{ $E_{\rm\,K\alpha} $ } & 
   \multicolumn{1}{c}{EW } & 
   \multicolumn{1}{c}{ $\log f_{\mathrm K\alpha}$} & 
   \multicolumn{1}{c}{ $\log L_{\rm\,K\alpha}$} &  \phantom{} &
   \multicolumn{1}{c}{p } & 
   \multicolumn{1}{c}{EW } &  \phantom{} &
   \multicolumn{1}{c}{ $f_{\rm\,obs}$ } &
   \multicolumn{1}{c}{$\log N_{\rm\,H}^{\mathrm{\,T,\,obs}}$ } &
   \multicolumn{1}{c}{ Ref.  }    \\ 

\noalign{\smallskip}

   \multicolumn{1}{l}{ } & 
   \multicolumn{1}{c}{  } & 
   \multicolumn{1}{c}{ } & 
   \multicolumn{1}{c}{[{\tiny $\rm\,erg\,cm^{-2}\,s^{-1}$}]  } & 
   \multicolumn{1}{c}{ [{\tiny $\rm\,erg\,s^{-1}$}]} &  \phantom{} &
   \multicolumn{1}{c}{[{\tiny keV}] } & 
   \multicolumn{1}{c}{[{\tiny eV}]} & 
      \multicolumn{1}{c}{ [{\tiny $\rm\,erg\,cm^{-2}\,s^{-1}$}] } & 
   \multicolumn{1}{c}{[{\tiny $\rm\,erg\,s^{-1}$}] } &  \phantom{} &
   \multicolumn{1}{c}{  } & 
   \multicolumn{1}{c}{ [{\tiny eV}] }  &  \phantom{} &
   \multicolumn{1}{c}{  } & 
   \multicolumn{1}{c}{ [{\tiny $\rm\,cm^{-2}$}]  } & 
   \multicolumn{1}{c}{  }    
   \\

\noalign{\smallskip}

\hline 
\noalign{\smallskip}

\endhead

\noalign{\smallskip}
\hline
\endfoot

\endlastfoot

\noalign{\smallskip} \hline \noalign{\smallskip}
B2 2201+31A  & A    &$1.79^{+0.04}_{-0.04}$ & $-11.38^{+0.01}_{-0.01}$ &  45.05  &     & $6.44^{+0.07}_{-0.08}$  & $53^{+18}_{-20}$ & $-13.4^{+0.2}_{-0.4}$ & 43.05  & & 13\% & --& &  0.09  &  23.76  & 1 \\ %
\noalign{\smallskip}
IC 4329A -- I&  E &$ 1.85^{+0.04 }_{-0.04 }$ &  $-9.763^{+0.006}_{-0.004}$&   44.00   &   &$6.41^{+0.03}_{-0.03}$  & $60^{+9}_{-9}$ & $-12.0^{+0.1 }_{-0.2 }$ & 41.76  & & 4\% & -- & &  0.24	 &  24.45 & 2\\ %
\noalign{\smallskip}
IC 4329A -- II&  G &$  1.71^{+0.01 }_{-0.02 }$ &  $-10.006^{+0.002}_{-0.003}$&   43.75  &   &$6.417^{+0.05}_{-0.013}$  & $43^{+3}_{-3}$ & $-12.29^{+0.06 }_{-0.06 }$ & 41.47  & & $>99\%$ & $59^{+5}_{-6}$& &  // & // &// \\ %
\noalign{\smallskip}
NGC 3227 -- I&  F	&$ 1.85^{+0.15}_{-0.17}$ & $-10.81^{+0.07 }_{-0.05 }$ & 41.67  &    & $6.39^{+0.01}_{-0.01}$& $185^{+15 }_{-7 }$  & $ -12.50^{+0.05 }_{-0.06}$  & 39.98  & &$\sim 0\%$ & -- & & 0.86	&  24.63  &2\\ %
\noalign{\smallskip}
NGC 3227 -- II& 	F &$ 1.63^{+ 0.01}_{-0.01 }$ & $-10.42^{+0.01 }_{-0.01 }$ &  42.06  &   & $6.41^{+0.01}_{-0.01}$  & $ 63^{+5 }_{-5}$ & $-12.46^{+ 0.06}_{-0.06 }$ & 40.02  & & $>99\%$ & $59^{+10}_{-10}$& & // & // & // \\ %
\noalign{\smallskip}
NGC 4151 -- I & C	&$ 1.56^{+0.03 }_{-0.03 }$ & $-10.118^{+0.004 }_{-0.004 }$ &   42.26  &  & $6.41^{+0.01}_{-0.01}$  & $ 205^{+7 }_{-6}$ & $-11.77^{+0.03 }_{-0.04 }$ & 40.61  & & $>99\%$ &  $125^{+18}_{-17}$& &  0.16&	  24.57  & 2 \\ %
\noalign{\smallskip}
NGC 4151 -- II & C	&$ 1.57^{+0.03 }_{-0.02 }$ & $-10.158^{+0.003 }_{-0.004 }$ & 42.22  &    & $6.41^{+0.01}_{-0.01}$  & $ 215^{+8}_{-7}$ & $-11.75^{+0.03 }_{-003 }$ & 40.63  & &$>99\%$ & $112^{+19}_{-17}$& & //  & // & // \\ %
\noalign{\smallskip}
NGC 4151 -- III&  C	&$ 1.50^{+0.02 }_{-0.03 }$ & $-10.159^{+0.006 }_{-0.01 }$ &  42.22  &   & $6.40^{+0.01}_{-0.01}$  & $ 207^{+8 }_{-7}$ & $-11.74^{+0.03 }_{-0.03 }$ & 40.64  & & $98\%$ & $111^{+17}_{-25}$ &  & //& // &//\\ %
\noalign{\smallskip}
NGC 4151 -- IV& C	&$ 1.63^{+0.02 }_{-0.03 }$ & $-9.499^{+0.007 }_{-0.007 }$ &  42.88  &   & $6.41^{+0.01}_{-0.01}$  & $68^{+4 }_{-3}$ & $-11.63^{+0.04 }_{-0.05 }$ & 40.75  & & $97\%$ & $33^{+17}_{-13}$ & & //& // & //\\ %
\noalign{\smallskip}
NGC 4151 -- V& C	&$ 1.63^{+0.04 }_{-0.02 }$ & $-9.509^{+0.008 }_{-0.008 }$ &  42.87  &   & $6.41^{+0.01}_{-0.01}$  & $ 73^{+3 }_{-4}$ & $-11.60^{+0.04 }_{-0.04 }$ & 40.78  & & $96\%$ & $36^{+14}_{-12}$&  & //& // & // \\ %
\noalign{\smallskip}
NGC 4151 -- VI&  C	&$ 1.57^{+ 0.06}_{-0.07 }$ & $-9.43^{+0.01 }_{-0.01 }$ & 42.95  &    & $6.41^{+0.01}_{-0.01}$  &  $58^{+3 }_{-3 }$ &$ -11.59^{+0.04 }_{-0.05}$ & 40.79  & &$>99\%$ & $136^{+30}_{-35}$& & // & // &// \\ %
\noalign{\smallskip}
NGC 4151 -- VII& C	&$1.73^{+0.04 }_{-0.04 }$ & $-10.03^{+0.02 }_{-0.02 }$ &  42.35  &   & $6.42^{+0.01}_{-0.01}$  & $ 208^{+7 }_{-8}$ & $-11.59^{+0.02 }_{-0.02 }$ & 40.79  & &$>99\%$ & $100^{+8}_{-7}$&  & // & // & // \\ %
\noalign{\smallskip}
NGC 4151 -- VIII& C	&$ 1.68^{+0.05 }_{-0.05 }$ & $-9.76^{+0.02 }_{-0.02 }$ &  42.62  &   & $6.43^{+0.01}_{-0.01}$  & $ 109^{+6 }_{-7}$ & $-11.68^{+0.03 }_{-0.04 }$ & 40.70  & &   $96\%$ & $68^{+23}_{-20}$&  & // & // & // \\ %
\noalign{\smallskip}
NGC 6814  &  B&$  1.88^{+0.01 }_{-0.01 }$ & $ -10.49^{+0.15 }_{-0.05 }$ &  42.29  &   & $6.4^{\mathrm{f}}$  & $82^{+17 }_{-15}$ & $-12.7 ^{+0.3 }_{-0.5 }$ & 40.08  &&   $91\%$ & --& & 0.13&	  23.67 & 3  \\ %
\noalign{\smallskip}
NGC 7469 -- I&  C	&$ 2.04^{+0.02 }_{-0.03 }$ & $-10.598^{+0.005 }_{-0.004 }$ &  43.16  &   & $6.40^{+0.02}_{-0.03}$  & $93^{+13 }_{-16}$ & $-12.41^{+0.08}_{-0.11 }$ &41.34  & & $\sim 0\%$ & -- & &  0.20&	  24.64  & 2 \\ %
\noalign{\smallskip}
NGC 7469 -- II&  C	&$2.11^{+0.02 }_{-0.02 }$ & $-10.37^{+0.09 }_{-0.03 }$ & 43.39  &    & $6.43^{+0.02}_{-0.02}$  & $75^{+16}_{-6}$ & $-12.53^{+0.12 }_{-0.16 }$ & 41.22  & & $\sim 0\%$ &  --& &// & //&// \\ %
\noalign{\smallskip}
NGC 7469 -- III&  C	&$ 2.09^{+0.01}_{-0.01 }$ & $-10.36^{+0.02 }_{-0.02 }$ &  43.40  &   & $6.42^{+0.01}_{-0.01}$   & $80^{+6 }_{-6 }$& $-12.46 ^{+0.05}_{-0.05}$ &  41.29  & & $\sim 0\%$ & --& &// &//  &// \\ %
\noalign{\smallskip}
NGC 7469 -- IV& 	F &$1.92^{+0.09 }_{-0.02 }$ & $-10.46^{+0.07 }_{-0.04 }$ &  43.29  &   & $6.44^{+0.02}_{-0.02}$  & $ 84 ^{+9}_{-5}$ & $-12.51^{+0.05 }_{-0.05 }$ & 41.24  & & $92\%$ & --& & // & // & //\\ %
\noalign{\smallskip}
PG 0050+124 -- I&   B  &$2.43^{+0.02}_{-0.03}$ & $-11.08^{+0.01}_{-0.01}$ & 43.88  &  & $6.45^{+0.05}_{-0.05}$ & $56^{+16}_{-19}$ & $-13.5^{+0.3}_{-1.8}$ & 41.45  & &$> 99\%$ & $162^{+65}_{-62}$& &  0.63  &  23.31  & 1\\ %
\noalign{\smallskip}
PG 0050+124 -- II &  C  &$2.24^{+0.02}_{-0.02}$ & $-11.333^{+0.003}_{-0.007}$& 43.63  & & $6.35^{+0.07}_{-0.03}$  & $29^{+15}_{-13}$ &$-13.9^{+0.3}_{-0.8}$  & 41.05  & & $> 99\%$ & $53^{+27}_{-18}$ & &//  & // &// \\ 
\noalign{\smallskip}
PG 0157+001 &  A   &$2.23^{+0.08}_{-0.09}$ & $-12.06^{+0.04}_{-0.04}$ & 43.80  & & $6.4^{\mathrm{f}}$  & $164^{+102}_{-100}$ & $-13.7^{+0.3}_{-1.0}$ & 42.16  & &56\% & -- & & 0.53  &  23.86 & 1  \\ %
\noalign{\smallskip}
PG 0838+770  &   A  &$1.52^{+0.06}_{-0.06}$ & $-12.24^{+0.04}_{-0.04}$ & 43.40  & & $6.43^{+0.07}_{-0.12}$  & $193^{+107}_{-84}$ & $-13.8^{+0.3}_{-0.7}$ & 41.96  & & $\sim0\%$ & --& & 0.48   & 23.71   & 1 \\ %
\noalign{\smallskip}
PG 0953+414  &  A   &$2.04^{+0.04}_{-0.04}$ & $-11.54^{+0.02}_{-0.02}$ &  44.70  &   & $6.4^{\mathrm{f}}$  & $10^{+13}_{-10}$ & $\leq -13.8$ & $\leq42.43$  & &$\sim 0\%$ & --& & 0.10   & 23.98 & 1  \\ %
\noalign{\smallskip}
PG 1004+130  &   B  &$1.57^{+0.16}_{-0.21}$ & $-12.45^{+0.06}_{-0.05}$ &  43.75  &   & $6.4^{\mathrm{f}}$  & $22^{+73}_{-22}$ & $\leq -14.20$ & $\leq 42.04$  & & $63\%$ & --& & 0.20  &  23.50 &  1 \\ %
\noalign{\smallskip}
PG 1116+215 -- I  &  A   &$2.29^{+0.04}_{-0.04}$ & $-11.51^{+0.02}_{-0.02}$ &  44.45  &    & $6.4^{\mathrm{f}}$& $7^{+36}_{-7}$ & $\leq -13.6$ & $\leq42.34$  & & $85\%$& -- & & 0.23   &  23.60 & 1 \\ %
\noalign{\smallskip}
PG 1116+215 -- II  &  C   &$2.37^{+0.02}_{-0.02}$ & $-11.207^{+0.002}_{-0.002}$&   44.76  &  & $6.46^{+0.07}_{-0.08}$  & $47^{+10}_{-10}$ & $-13.7^{+0.2}_{-0.4}$  & 42.24  & & $\sim 0\%$ & --& &// &// & //    \\ %
\noalign{\smallskip}
PG 1116+215 -- III   &  A   &$2.12^{+0.06}_{-0.06}$ & $-11.46^{+0.02}_{-0.02}$ &  44.48  &   & $6.4^{\mathrm{f}}$  & $76^{+37}_{-36}$ & $\leq -13.4$ &$\leq 42.59$  & &$88\%$ & --& &// & //& //    \\ 
\noalign{\smallskip}
PG 1116+215 -- IV  &  C   &$ 2.39^{+0.03}_{-0.03}$ & $ -11.08^{+0.05}_{-0.06}$ &  44.88  &   & $6.4^{\mathrm{f}}$  & $27^{+9}_{-10}$ & $\leq -13.5$ & $\leq 42.44$  & &$\sim 0\%$  & --& & //& // &// \\ %
\noalign{\smallskip}
PG 1116+215 -- V   &   C  &$ 2.16^{+0.03}_{-0.03}$ & $ -11.147^{+0.002}_{-0.002}$ & 44.80  &   & $6.39^{+0.11}_{-0.12}$  & $20^{+9}_{-9}$ & $-13.8^{+0.3}_{-0.8}$ & 42.14  & &$\sim 0\%$ & --&  &// & // &// \\ %
\noalign{\smallskip}
PG 1116+215 -- VI  &  C  &$2.33^{+0.03}_{-0.04}$ & $-11.310^{+0.002}_{-0.002}$ & 44.65  &   & $6.40^{+0.13}_{-0.08}$  & $33^{+12}_{-13}$ & $-13.7^{+0.2}_{-0.4}$ & 42.24  & &$\sim 0\%$ & -- & & // & // & // \\ %
\noalign{\smallskip}
PG 1126$-$041 -- I   &  D   &$2.52^{+0.10}_{-0.05}$ & $-11.74^{+0.03}_{-0.03}$&   43.20  &  & $6.4^{\mathrm{f}}$  & $97^{+31}_{-33}$ & $-13.8^{+0.2}_{-0.6}$  & 41.14  &  & $51\%$ & --& & 0.37  &  23.88 & 1  \\ %
\noalign{\smallskip}
PG 1126$-$041 -- II &    D &$2.8^{+0.2}_{-0.4}$ & $-11.39^{+0.06}_{-0.06}$ &  43.56  &   & $6.48^{+0.03}_{-0.12}$  & $95^{+71}_{-49}$ & $\leq -13.4$  & $\leq 41.54$  & &$55\%$& --& &// &// & //  \\%
\noalign{\smallskip}
PG 1126$-$041 -- III &   D  &$2.5^{+0.3}_{-0.3}$ & $-11.37^{+0.04}_{-0.09}$ &   43.59  &  & $6.4^{\mathrm{f}}$  & $24^{+74}_{-24}$ & $\leq -13.5$ & $\leq 41.44$  & &$21\%$ & -- & &// &// & // \\
\noalign{\smallskip}
PG 1126$-$041 -- IV &  D   &$2.68^{+0.07}_{-0.05}$ & $-11.60^{+0.01}_{-0.04}$&  43.35  &   & $6.47^{+0.02}_{-0.03}$   & $100^{+84}_{-4}$ & $-13.2^{+0.2}_{-0.2}$ & 41.74  & & $98\%$  & $230^{+48}_{-73}$& &// &//   & // \\ %
\noalign{\smallskip}
PG 1229+204  &   A  &$2.03^{+0.03}_{-0.04}$ & $-11.51^{+0.01}_{-0.01}$ &  43.48  &   & $6.40^{+0.08}_{-0.08}$  & $98^{+28}_{-24}$ & $-13.5^{+0.2}_{-0.3}$ & 41.49  &  & 35\%& -- & & 0.31  &  23.97 & 1 \\ %
\noalign{\smallskip}
PG 1244+026  &   A  &$2.17^{+0.09}_{-0.09}$ & $-11.67^{+0.02}_{-0.03}$ &   43.07  &  & $6.4^{\mathrm{f}}$  & $25^{+66}_{-25}$ & $\leq -13.4$ & $\leq 41.34$  & & 42\% & --& & 0.33  &  23.93 & 1 \\ %
\noalign{\smallskip}
PG 1309+355  &   A  &$1.68^{+0.04}_{-0.03}$ & $-12.15^{+0.02}_{-0.02}$ &  43.80  &   & $6.37^{+0.11}_{-0.08}$  & $107^{+44}_{-43}$ & $-13.9^{+0.2}_{-0.4}$ & 42.07  & & 44\% & -- & & 0.16   & 23.92  & 1 \\ %
\noalign{\smallskip}
PG 1411+442  &  C   &$1.66^{+0.12}_{-0.11}$ & $-12.09^{+0.07}_{-0.09}$ &  43.20  &    & $6.4^{\mathrm{f}}$& $154^{+72}_{-61}$  & $\leq -13.5$ &$\leq 41.80$  & & $>99$\% & $669_{-528}^{+456}$& &  0.48  &  23.46  & 1 \\ %
\noalign{\smallskip}
PG 1426+015  &  A   &$2.04^{+0.05}_{-0.05}$ & $-11.09^{+0.02}_{-0.02}$ &  44.18  &   & $6.3^{+0.4}_{-0.1}$ & $83^{+40}_{-42}$ & $-13.2^{+0.3}_{-1.1}$ & 42.06  & & $\sim 0\%$ & --& & 0.33  &  23.89 &  1  \\ %
\noalign{\smallskip}
PG 1435$-$067   &   A  &$2.06^{+0.05}_{-0.06}$ & $-11.62^{+0.02}_{-0.02}$ &  44.03   &  & $6.4^{\mathrm{f}}$  & $100^{+35}_{-41}$ & $-13.7^{+0.3}_{-0.9}$ &41.94  &  & 65\% & -- & &  0.32 &   23.17 & 1  \\ %
\noalign{\smallskip}
PG 1440+356 -- I &  A   &$2.39^{+0.03}_{-0.03}$ & $-11.62^{+0.01}_{-0.01}$ &  43.56  &   & $6.41^{+0.15}_{-0.20}$  & $69^{+25}_{-27}$ & $\leq -13.7$ & $\leq 41.46$  & & 86\% & --& & 0.36 &   23.53 &  1 \\ %
\noalign{\smallskip}
PG 1440+356 -- II &  A   &$2.32^{+0.03}_{-0.03}$ & $-11.52^{+0.01}_{-0.01}$ &  43.65  &   & $6.4^{\mathrm{f}}$  & $40^{+26}_{-22}$ & $\leq -13.6$ & $\leq 41.56$  & &88\%  &--&  &// & //& //  \\ %
\noalign{\smallskip}
PG 1440+356 -- III & A    &$2.38^{+0.04}_{-0.04}$ & $-11.64^{+0.02}_{-0.02}$ &  43.54  &   & $6.33^{+0.08}_{-0.11}$  & $81^{+27}_{-39}$ & $\leq -13.7$ & $\leq 41.46$  & & 71\% & --& &// & //&  // \\ %
\noalign{\smallskip}
PG 1440+356 -- IV &   A  &$2.28^{+0.04}_{-0.04}$ & $-11.76^{+0.02}_{-0.02}$ &  43.41  &   & $6.4^{\mathrm{f}}$  & $56^{+33}_{-44}$ & $\leq -13.7$ & $\leq 41.46$  & & 80\% & -- & &// & //&//  \\ %
\noalign{\smallskip}
PG 1448+273  &  A   &$2.29^{+0.04}_{-0.04}$ & $-11.71^{+0.01}_{-0.01}$ &   43.30  &  & $6.42^{+0.06}_{-0.07}$  & $118^{+32}_{-34}$ & $-13.7^{+0.2}_{-0.4}$ & 41.30  & & 61\% & --& & 0.15  &  23.99 &  1  \\ %
\noalign{\smallskip}
PG 1613+658 -- I &  A   &$1.76^{+0.06}_{-0.06}$ & $-11.40^{+0.03}_{-0.03}$ & 44.29   &   & $6.32^{+0.10}_{-0.04}$  & $54^{+75}_{-54}$ & $\leq -13.1$ & $\leq 42.61$  & & 5\% & --  & &  0.45  &  23.69& 1 \\ %
\noalign{\smallskip}
PG 1613+658 -- II &  A   &$1.89^{+0.07}_{-0.07}$ & $-11.31^{+0.03}_{-0.03}$ &  44.39  &   & $6.42^{+0.09}_{-0.10}$  & $132^{+68}_{-64}$ & $\leq -12.9$ & $\leq 42.81$  & & 5\% & --& &// & //& //  \\ %
\noalign{\smallskip}
PG 1626+554  &   A  &$2.04^{+0.06}_{-0.07}$ & $-11.55^{+0.03}_{-0.03}$ &  44.12  &   & $6.47^{+0.06}_{-0.12}$  & $144^{+62}_{-66}$ & $-13.3^{+0.3}_{-1.0}$ & 42.36  & & 4\% & -- & & 0.18  &  23.32   & 1\\%
\noalign{\smallskip}
PG 2214+139  &  C  &$2.12^{+0.07}_{-0.07}$ & $-11.32^{+0.02}_{-0.01}$ &  43.71  &   & $6.35^{+0.11}_{-0.07}$  & $71^{+36}_{-24}$ & $-13.4^{+0.2}_{-0.4}$ & 41.63  & & 6\%& -- & & 0.13  &  23.77  & 1 \\ %

\noalign{\smallskip}
\hline
\noalign{\smallskip}
\multicolumn{17}{l}{{\bf Notes.} $^f$: energy of the line fixed to 6.4\,keV. The values of the torus parameters were taken from: (1) \citet{Mor:2009fk}, (2) \citet{Alonso-Herrero:2011zr}, (3) \citet{Ramos-Almeida:2011ly}. } \\
\multicolumn{17}{l}{Typical uncertainties on the torus parameters are about 30\%.} \\
\end{longtable}
\end{center}
\normalsize
\end{landscape}
}

\clearpage

\longtab{3}{
\begin{landscape}
\begin{center}
\LTcapwidth=1.13\textwidth
\begin{longtable}{lccccccccccc}

\caption{Results of the spectral fits of the {\it XMM-Newton} observations of our sample. (1) Model used for the fit, (2) temperature of the bremsstrahlung component, (3 and 6) ionisation parameters, (4 and 7) column densities and (5 and 8) covering factors of the warm absorbers, (9) column density of the additional neutral absorber, (10) presence of additional emission lines (for details see Sect.\,\ref{Sect:spectral_analysis} and Appendix\,\ref{Appendix1}), and (11) chi-squared and degrees of freedom of the fit.}\label{tab:fitresults2}\\
\hline \hline \\


   \multicolumn{1}{l}{ } & 
   \multicolumn{1}{c}{(1) } & 
   \multicolumn{1}{c}{(2) } & 
   \multicolumn{1}{c}{(3) } & 
   \multicolumn{1}{c}{ (4)} & 
   \multicolumn{1}{c}{(5) } & 
   \multicolumn{1}{c}{(6) } & 
      \multicolumn{1}{c}{ (7)} & 
   \multicolumn{1}{c}{(8) } & 
   \multicolumn{1}{c}{(9) } & 
   \multicolumn{1}{c}{(10) }  &
   \multicolumn{1}{c}{(11) }  \\
\noalign{\smallskip}

   \multicolumn{1}{l}{ Source } & 
   \multicolumn{1}{c}{Model } & 
   \multicolumn{1}{c}{ $kT$} & 
   \multicolumn{1}{c}{$\log \xi^1$ } & 
   \multicolumn{1}{c}{$N_{\rm\,H}^{\rm\,W,\,1}$ } &  
   \multicolumn{1}{c}{ $f^{\,1}$ } & 
   \multicolumn{1}{c}{$\log \xi^2$  } & 
   \multicolumn{1}{c}{ $N_{\rm\,H}^{\rm\,W,\,2}$} & 
   \multicolumn{1}{c}{ $f^{\,2}$} & 
   \multicolumn{1}{c}{$N_{\rm\,H}^{\rm\,C}$ } & 
   \multicolumn{1}{c}{Lines } &
   \multicolumn{1}{c}{ $\chi^{2}$/DOF } \\ 

\noalign{\smallskip}

   \multicolumn{1}{l}{ } & 
   \multicolumn{1}{c}{  } & 
   \multicolumn{1}{c}{ [{\tiny eV}] } & 
   \multicolumn{1}{c}{[{\tiny $\rm\,erg\,cm\,s^{-1}$}]  } & 
   \multicolumn{1}{c}{ [{\tiny $10^{\,22}\rm\,cm^{-2}$}]} &  
   \multicolumn{1}{c}{ } & 
   \multicolumn{1}{c}{[{\tiny $\rm\,erg\,cm\,s^{-1}$}] } & 
      \multicolumn{1}{c}{ [{\tiny $10^{\,22}\rm\,cm^{-2}$}] } & 
   \multicolumn{1}{c}{ } &  
   \multicolumn{1}{c}{ [{\tiny $10^{\,22}\rm\,cm^{-2}$}] } & 
   \multicolumn{1}{c}{  }  &
      \multicolumn{1}{c}{  }  \\
\noalign{\smallskip}

\noalign{\smallskip}

\endfirsthead

\multicolumn{11}{l}%
{\small \hspace{-0.1in}{{\textbf{\tablename\ \thetable{}.} Results of the spectral fits of the {\it XMM-Newton} observations of our sample. -- \textit{continued} }}} \\
\noalign{\smallskip}
\hline \hline \\[0.05ex]


   \multicolumn{1}{l}{ } & 
   \multicolumn{1}{c}{(1) } & 
   \multicolumn{1}{c}{(2) } & 
   \multicolumn{1}{c}{(3) } & 
   \multicolumn{1}{c}{ (4)} & 
   \multicolumn{1}{c}{(5) } & 
   \multicolumn{1}{c}{(6) } & 
      \multicolumn{1}{c}{ (7)} & 
   \multicolumn{1}{c}{(8) } &  
   \multicolumn{1}{c}{(9) } & 
   \multicolumn{1}{c}{(10) }  &
   \multicolumn{1}{c}{(11) }  \\
\noalign{\smallskip}

   \multicolumn{1}{l}{ Source } & 
   \multicolumn{1}{c}{Model } & 
   \multicolumn{1}{c}{ $kT$} & 
   \multicolumn{1}{c}{$\log \xi^1$ } & 
   \multicolumn{1}{c}{$N_{\rm\,H}^{\rm\,W,\,1}$ } &  
   \multicolumn{1}{c}{ $f^{\,1}$ } & 
   \multicolumn{1}{c}{$\log \xi^2$  } & 
   \multicolumn{1}{c}{ $N_{\rm\,H}^{\rm\,W,\,2}$} & 
   \multicolumn{1}{c}{ $f^{\,2}$} & 
   \multicolumn{1}{c}{$N_{\rm\,H}^{\rm\,C}$ } & 
   \multicolumn{1}{c}{Lines } & 
   \multicolumn{1}{c}{$\chi^{2}$/DOF } \\ 

\noalign{\smallskip}

   \multicolumn{1}{l}{ } & 
   \multicolumn{1}{c}{  } & 
   \multicolumn{1}{c}{ [{\tiny eV}] } & 
   \multicolumn{1}{c}{[{\tiny $\rm\,erg\,cm\,s^{-1}$}]  } & 
   \multicolumn{1}{c}{ [{\tiny $\rm\,cm^{-2}$}]} &  
   \multicolumn{1}{c}{ } & 
   \multicolumn{1}{c}{[{\tiny $\rm\,erg\,cm\,s^{-1}$}] } & 
      \multicolumn{1}{c}{ [{\tiny $10^{\,22}\rm\,cm^{-2}$}] } & 
   \multicolumn{1}{c}{ } &  
   \multicolumn{1}{c}{ [{\tiny $10^{\,22}\rm\,cm^{-2}$}] } & 
   \multicolumn{1}{c}{  }  &
   \multicolumn{1}{c}{  }  \\
\noalign{\smallskip}

\hline 
\noalign{\smallskip}

\endhead

\noalign{\smallskip}
\hline
\endfoot

\endlastfoot

\noalign{\smallskip} \hline \noalign{\smallskip}
B2 2201+31A  & A    & $263^{+28}_{-28}$  & --  & --& --& --& --& --& --& --& 969.1/927 \\%
\noalign{\smallskip}
IC 4329A -- I&  E & $439^{+42}_{-33}$ & $-0.17^{+0.25}_{-0.27}$ & $1.04^{+0.11}_{-0.07}$& $0.87^{+0.02}_{-0.02}$ & -- & --& --&  $0.22^{+0.01}_{-0.01}$ & -- &2281.6/2079\\%
\noalign{\smallskip}
IC 4329A -- II&  G & $594^{+24}_{-22}$  & $-0.30^{+0.06}_{-0.05}$ & $0.72^{+0.03}_{-0.03}$ & $0.79^{+0.01}_{-0.01} $ & $2.43^{+0.05}_{-0.05}$ & $0.26^{+0.26}_{-0.06}$ & $0.88^{+0.03}_{-0.40}$& $0.20^{+0.01}_{-0.01}$ & $\checkmark$ & 3810.1/3136\\%
\noalign{\smallskip}
NGC 3227 -- I&  F	& $565^{+83}_{-69}$ & $-0.1^{+0.3}_{-0.2}$ &$7.8^{+0.2}_{-0.3}$ & $> 0.99$ & $-1.2^{+1.2}_{-0.5}$& $0.14^{+0.03}_{-0.05} $ & $0.7^{+0.2}_{-0.2}$ & -- & -- &1781.5/1716\\%
\noalign{\smallskip}
NGC 3227 -- II& 	F  & $657^{+30}_{-10}$  & $2.10^{+0.02}_{-0.02}$ &$3.00^{+0.40}_{-0.02}$ & $0.68^{+0.01}_{-0.01}$&$-0.67^{+0.02}_{-0.03}$ & $0.22^{+0.01}_{-0.02}$ & $0.88^{+0.01}_{-0.01}$& -- & -- &3603.3/3032 \\%
\noalign{\smallskip}
NGC 4151 -- I & C	& $508^{+7}_{-14}$ &$0.54^{+0.13}_{-0.14}$  & $8.08^{+0.05}_{-0.06}$ & $0.992^{+0.002}_{-0.002}$& -- & --& --& -- & \checkmark & 3321.5/2808 \\%
\noalign{\smallskip}
NGC 4151 -- II & C	& $493^{+14}_{-24}$ & $0.41^{+0.11}_{-0.08}$ &$8.02^{+0.05}_{-0.05}$ & $0.994^{+0.002}_{-0.002}$ & -- & --& --& --&\checkmark & 3120.4/2785 \\%
\noalign{\smallskip}
NGC 4151 -- III&  C	& $506^{+14}_{-11}$ & $0.43^{+0.14}_{-0.06}$ & $8.01^{+0.07}_{-0.04}$ & $0.993^{+0.001}_{-0.002}$ & -- & --& --& --&\checkmark & 2839.0/2683 \\%
\noalign{\smallskip}
NGC 4151 -- IV& C	& $801^{+37}_{-48}$ & $0.82^{+0.03}_{-0.13}$ & $7.60^{+0.03}_{-0.03}$ & $0.992^{+0.001}_{-0.001}$ & -- & --& --& --&\checkmark &3304.3/2988\\%
\noalign{\smallskip}
NGC 4151 -- V& C	& $842^{+26}_{-54}$  &$0.76^{+0.10}_{-0.16}$  & $7.37^{+0.06}_{-0.10}$ & $0.990^{+0.002}_{-0.001}$  & -- & --& --& --&\checkmark &3281.8/2967\\%
\noalign{\smallskip}
NGC 4151 -- VI&  C	&$1040^{+66}_{-84}$  & $0.64^{+0.07}_{-0.08}$ & $7.04^{+0.08}_{-0.09}$& $0.993^{+0.001}_{-0.001}$ & -- & --& --& --&\checkmark &3257.2/3016 \\%
\noalign{\smallskip}
NGC 4151 -- VII& C	& $360^{+132}_{-53}$ & $1.10^{+0.01}_{-0.01	}$ & $16.9^{+0.3}_{-0.3}$ & $0.979^{+0.002}_{-0.002}$& -- & --& --& --&\checkmark &3306.5/2924   \\%
\noalign{\smallskip}
NGC 4151 -- VIII& C	&$286^{+17}_{-16}$  & $1.10^{+0.06}_{-0.06}$ & $12.8^{+0.3}_{-0.5}$& $0.984^{+0.002}_{-0.002}$ & -- & --& --& --&\checkmark & 3420.0/3130 \\%
\noalign{\smallskip}
NGC 6814  &  B&  -- &$2.20^{+0.11}_{-0.06}$ & $65^{+45}_{-3}$ & $ 0.42^{+0.14}_{-0.05}$ & -- & --& --& --& --&1592.4/1525 \\%
\noalign{\smallskip}
NGC 7469 -- I&  C	& $233^{+34}_{-34}$ &$2.21^{+0.08}_{-0.05}$  & $66^{+31}_{-4}$ & $0.45^{+0.11}_{-0.05}$&-- &-- & --& --& --& 1954.6/1870 \\%
\noalign{\smallskip}
NGC 7469 -- II&  C	&$222^{+19}_{-22}$  & $2.25^{+0.06}_{-0.05}$ & $65^{+25}_{-2}$ & $0.44^{+0.08}_{-0.04}$ & -- &--  &--  &--  &--  & 2323.4/2088 \\%
\noalign{\smallskip}
NGC 7469 -- III&  C	& $225^{+10}_{-9}$ &  $2.18^{+0.02}_{-0.03}$ &$65^{+7}_{-2}$ & $0.44^{+0.03}_{-0.03}$ &--  &--  &--  & -- & --  &1897.1/1768 \\%
\noalign{\smallskip}
NGC 7469 -- IV& 	F & $515^{+11}_{-11}$ & $-1.10^{+0.12}_{-0.12}$ & $0.84^{+0.07}_{-0.11}$ & $0.75^{+0.02}_{-0.05}$ & $2.7^{+0.3}_{-0.5}$&$45^{+16}_{-5}$ & $0.14^{+0.07}_{-0.01}$ & -- & -- & 1876.3/1759\\%
\noalign{\smallskip}
PG 0050+124 -- I&   B  & -- &$-1.10^{+0.11}_{-0.11}$  &  $0.06^{+0.01}_{-0.01}$ & $\geq 0.91$ & --&-- &-- &--  & \checkmark &1510.0/1394 \\%
\noalign{\smallskip}
PG 0050+124 -- II &  C  & $322^{+13}_{-21}$  &$-0.23_{-0.11}^{+0.06}$  & $0.07^{+0.02}_{-0.01}$& $\geq 0.88$ & --&-- &-- &--  & \checkmark& 1927.1/1910 \\%
\noalign{\smallskip}
PG 0157+001 &  A   & $192^{+53}_{-50}$ &--  &-- &-- &-- &-- & --& --& --& 364.6/351 \\%
\noalign{\smallskip}
PG 0838+770  &   A  &$164^{+17}_{-16}$  & -- &-- &-- &-- & --&-- &-- &-- & 349.6/305 \\%
\noalign{\smallskip}
PG 0953+414  &  A   & $288^{+17}_{-17}$  &--  &-- & --&-- &-- & --& --& --&861.9/827 \\%
\noalign{\smallskip}
PG 1004+130  &   B  & -- & $2.1^{+1.7}_{-1.4}$ &$0.6^{+0.8}_{-0.4}$ &$0.5^{+0.2}_{-0.2}$ &-- & --&-- &-- &-- & 127.4/117 \\%
\noalign{\smallskip}
PG 1116+215 -- I  &  A   & $91^{+4}_{-4}$  &  --& --& --& --&-- & --& --&  \checkmark & 743.0/665\\%
\noalign{\smallskip}
PG 1116+215 -- II  &  C   & $193^{+12}_{-12}$ & $0.8^{+0.4}_{-0.7}$ & $40^{+7}_{-18}$& $0.60^{+0.09}_{-0.04}$ & --&-- &-- & --& --&2032.7/1826 \\%
\noalign{\smallskip}
PG 1116+215 -- III   &  A   & $269^{+26}_{-26}$ & -- & --& --&-- & --& --&-- &\checkmark &713.5/668 \\%
\noalign{\smallskip}
PG 1116+215 -- IV  &  C   &$192^{+13}_{-14}$  & $-0.2^{+0.3}_{-0.3}$ & $18^{+4}_{-3}$& $0.48^{+0.03}_{-0.03}$ & --&-- & --&-- & -- & 2216.4/2060 \\%
\noalign{\smallskip}
PG 1116+215 -- V   &   C  & $194^{+10}_{-10}$  & $1.4^{+0.4}_{-0.2}$  & $35^{+9}_{-8}$& $0.48^{+0.02}_{-0.05}$ &-- &-- &-- &-- &-- & 2214.2/2125 \\%
\noalign{\smallskip}
PG 1116+215 -- VI  &  C &$176^{+13}_{-13}$  & $-0.2^{+0.3}_{-0.3}$ &$18^{+3}_{-3}$ & $0.57^{+0.03}_{-0.03}$ & --& --& --&-- &-- & 2128.5/1919\\%
\noalign{\smallskip}
PG 1126$-$041 -- I   &  D  & -- & $0.37^{+0.12}_{-0.07}$  & $6.3^{+0.1}_{-0.3}$ & $0.986^{+0.001}_{-0.001}$ & $3.33^{+0.05}_{-0.10}$ & $58^{+30}_{-18}$ & $\geq 0.86$ &-- &\checkmark & 383.9/381 \\%
\noalign{\smallskip}
PG 1126$-$041 -- II &    D &--  & $-1.3^{+1.5}_{-0.1}$ &$1.7^{+1.4}_{-0.6} $ & $0.98^{+0.01}_{-0.02}$ & $2.71^{+0.2}_{-0.5}$ &$58^{+11}_{-10}$ & $0.78^{+0.15}_{-0.12}$ & -- & \checkmark & 268.0/284 \\%
\noalign{\smallskip}
PG 1126$-$041 -- III &   D  & -- & $0.6^{+0.4}_{-0.2}$ & $3.7^{+1.7}_{-1.0}$ & $0.98^{+0.01}_{-0.02}$ &$2.7^{+0.4}_{-0.2}$ & $3^{+3}_{-2}$ &$\geq 0.72$ & -- &\checkmark & 168.2/167\\%
\noalign{\smallskip} 
PG 1126$-$041 -- IV &  D   & -- & $-1.06^{+0.11}_{-0.04}$ & $2.4^{+0.1}_{-0.2}$ & $0.981^{+0.002}_{-0.001}$& $2.73^{+0.01}_{-0.03}$ & $59^{+3}_{-3}$ &$0.73^{+0.02}_{-0.02}$ & --&\checkmark & 1064.0/953 \\%
\noalign{\smallskip}
PG 1229+204  &   A  & $268^{+19}_{-19}$ & -- & --&-- &-- & --& --& --& \checkmark & 1120.6/1005 \\%
\noalign{\smallskip}
PG 1244+026  &   A  & $408^{+13}_{-13}$ &  --&-- &-- &-- &-- &-- &-- &-- & 624.0/610\\%
\noalign{\smallskip}
PG 1309+355  &   A  & $148^{+14}_{-14}$ & -- & --&-- &-- &-- &-- & --& --& 521.2/495\\%
\noalign{\smallskip}
PG 1411+442  &  C   & $261^{+52}_{-28}$ &$-0.8^{+0.6}_{-0.1}$  &$7^{+20}_{-4}$ &$0.90^{+0.08}_{-0.09}$ & --& --& --& --&-- & 162.7/140\\%
\noalign{\smallskip}
PG 1426+015  &  A   &   $258^{+22}_{-22}$ &-- &-- &-- & --& --& --& --&-- & $734.8/692$ \\%
\noalign{\smallskip}
PG 1435$-$067   &   A & $267^{+24}_{-24}$ & -- &-- &-- &-- &-- & --&-- &-- & 858.5/851 \\%
\noalign{\smallskip}
PG 1440+356 -- I &  A   & $238^{+6}_{-6}$  & -- &-- &-- & --&-- & --& --& --& 1218.2/1148 \\%
\noalign{\smallskip}
PG 1440+356 -- II &  A  & $222^{+7}_{-7}$ & -- &-- &-- &-- & --& --& --&\checkmark & 1187.6/1081\\%
\noalign{\smallskip}
PG 1440+356 -- III & A   & $223^{+8}_{-8}$ & -- & --& --&-- & --&-- &-- &\checkmark & 1259.1/1117 \\%
\noalign{\smallskip}
PG 1440+356 -- IV &   A  & $224^{+9}_{-9}$  &  --&-- & --& --& --& --&-- & \checkmark &883.7/874 \\%
\noalign{\smallskip}
PG 1448+273  &  A   & $302^{+9}_{-9}$ &--  & --&-- &-- &-- &-- &-- &-- & 941.9/897 \\%
\noalign{\smallskip}
PG 1613+658 -- I &  A   & $173^{+26}_{-26}$ &  --&-- &-- & --&-- &-- &-- & -- & 302.3/303\\%
\noalign{\smallskip}
PG 1613+658 -- II &  A   & $234^{+32}_{-32}$ & -- &-- &-- &-- &-- & --&-- &-- & 431.4/461\\%
\noalign{\smallskip}
PG 1626+554  &   A  &$204^{+46}_{-46}$  & -- &-- & --& --& --&-- &-- &-- &591.1/586 \\%
\noalign{\smallskip}
PG 2214+139  &  C  & $317^{+20}_{-20}$ & $-1.3^{+0.1}_{-0.2}$  &$2.7^{+0.2}_{-0.2}$ &$0.997^{+0.001}_{-0.002}$ & --&-- &-- &-- &-- & 847.1/766 \\%
\noalign{\smallskip}
\hline
\noalign{\smallskip}
\end{longtable}
\end{center}
\normalsize
\end{landscape}
}

\appendix
\section{X-ray spectral results}\label{Appendix0}
In Tables\,\ref{tab:fitresults} and \ref{tab:fitresults2} we report the results of the spectral fitting described in Sect.\,\ref{Sect:spectral_analysis} for our {\it XMM-Newton}/EPIC sample.

\section{Notes on the individual sources}\label{Appendix1}
In the following we report the details on the emission lines added to the best fits.
\smallskip\newline
{\bf IC 4329a -- II.} A narrow ($\sigma=1$\,eV) emission feature is detected in the spectrum. The narrow line is found at $E=7.00^{+0.02}_{-0.03}$\,keV ($EW=20^{+2}_{-3}$\,eV), and is likely to be Fe\,XXVI. 
\smallskip\newline
 {\bf NGC 4151 -- I.}  The spectrum shows evidence of three narrow emission features at low energies. These features were fitted using Gaussian emission lines at $E=0.561^{+0.003}_{-0.003}$\,keV ($EW=91^{+3}_{-4}$\,eV), $E=0.903^{+0.003}_{-0.004}$\,keV ($EW=83^{+4}_{-3}$\,eV) and $E=1.33^{+0.01}_{-0.01}$\,keV ($EW=37^{+3}_{-6}$\,eV). These lines are consistent with O\,VII, Ne\,IX, and Mg\,XI, respectively. 
\smallskip\newline
 {\bf NGC 4151 -- II.} Three narrow emission features at low energies are detected. The narrow lines were found to be at $E=0.557^{+0.004}_{-0.002}$\,keV ($EW=89^{+4}_{-2}$\,eV), $E=0.903^{+0.003}_{-0.004}$\,keV ($EW=78^{+4}_{-3}$\,eV), and $E=1.34^{+0.01}_{-0.01}$\,keV ($EW=33^{+3}_{-4}$\,eV), and they are consistent with O\,VII, Ne\,IX, and Mg\,XI, respectively. 
 \smallskip\newline
{\bf NGC 4151 -- III.} Three narrow emission features at low energies are detected. The three lines are at $E=0.557^{+0.004}_{-0.003}$\,keV ($EW=78^{+3}_{-4}$\,eV), $E=0.897^{+0.007}_{-0.005}$\,keV ($EW=75^{+4}_{-3}$\,eV), and $E=1.31^{+0.02}_{-0.02}$\,keV ($EW=37^{+4}_{-6}$\,eV), and they are consistent with O\,VII, Ne\,IX, and Mg\,XI, respectively. 
\smallskip\newline
{\bf NGC 4151 -- IV.} Two narrow emission features at low energies are detected.  The narrow features are found at $E=0.558^{+0.003}_{-0.003}$\,keV ($EW=103^{+13}_{-1}$\,eV), and $E=0.897^{+0.007}_{-0.005}$\,keV ($EW=76^{+4}_{-5}$\,eV), and they are consistent with O\,VII and Ne\,IX, respectively.
\smallskip\newline
{\bf NGC 4151 -- V.} Two narrow emission features at low energies are detected. The narrow features are located at $E=0.558^{+0.003}_{-0.002}$\,keV ($EW=108^{+7}_{-8}$\,eV), and $E=0.892^{+0.009}_{-0.007}$\,keV ($EW=63^{+4}_{-4}$\,eV), and they are consistent with O\,VII and Ne\,IX, respectively. 
\smallskip\newline
{\bf NGC 4151 -- VI.} 
Two narrow emission features at $E=0.559^{+0.003}_{-0.002}$\,keV ($EW=100^{+13}_{-7}$\,eV) and $E=0.895^{+0.008}_{-0.005}$\,keV ($EW=71^{+4}_{-3}$\,eV) are detected. The two lines are consistent with O\,VII and Ne\,IX, respectively. 
\smallskip\newline
{\bf NGC 4151 -- VII.} Five narrow emission features at low energies are detected. The energies of the narrow lines are $E=0.563^{+0.002}_{-0.002}$\,keV ($EW=86^{+4}_{-3}$\,eV), $E=0.896^{+0.004}_{-0.003}$\,keV ($EW=77^{+3}_{-3}$\,eV), $E=1.34^{+0.01}_{-0.01}$\,keV ($EW=35^{+3}_{-3}$\,eV), $E=1.82^{+0.02}_{-0.02}$\,keV ($EW=69^{+4}_{-5}$\,eV), and $E=7.04^{+0.03}_{-0.07}$\,keV ($EW= 32^{+5}_{-4}$\,eV).
These lines are consistent with being due to O\,VII, Ne\,IX, Mg\,XI, Si\,XIII, and to Fe\,XXVI, respectively. 
\smallskip\newline
 {\bf NGC 4151 -- VIII.} Three narrow emission features at low energies are detected. The narrow lines have energies of $E=0.596^{+0.005}_{-0.005}$\,keV ($EW=77^{+3}_{-2}$\,eV), $E=0.897^{+0.006}_{-0.006}$\,keV ($EW=59^{+4}_{-3}$\,eV), and $E=1.78^{+0.04}_{-0.02}$\,keV ($EW=50^{+5}_{-4}$\,eV), and are consistent with O\,VII, Ne\,IX, and Si\,XIII, respectively. 
 \smallskip\newline
{\bf PG 0050+124 - I.} The spectrum requires an additional line at $E=6.96_{-0.10}^{+0.09}$ keV, likely due to Fe\,XXVI, with an equivalent width $EW=106\pm24\rm\,eV$. \smallskip\newline
{\bf PG 0050+124 - II.} Besides the Fe\,XXVI line at $E=6.97^{+0.06}_{-0.04}$ keV ($EW=56\pm13\rm\,eV$), we found evidence of another unresolved ionised iron line (likely Fe\,XXV) at $E=6.66^{+0.05}_{-0.04}$ keV ($EW=50_{-10}^{+12}\rm\,eV$). 
\smallskip\newline
{\bf PG 1116+215 -- I.} Two lines at $6.7\rm\,keV$ ($EW=63^{+38}_{-34}\rm\,eV$) and $6.97\rm\,keV$ ($EW=128^{+49}_{-45}\rm\,eV$) are needed. The lines are consistent with being produced by emission of ionised iron (Fe\,XXV and Fe\,XXVI, respectively). 
\smallskip\newline
{\bf PG 1116+215 -- III.} Two lines at $6.7\rm\,keV$ ($EW=68^{+35}_{-34}\rm\,eV$) and $6.97\rm\,keV$ ($EW=124^{+51}_{-45}\rm\,eV$) are needed. The two lines are consistent with being produced by emission of ionised iron (Fe\,XXV and Fe\,XXVI, respectively). 
\smallskip\newline
{\bf PG 1126$-$041 -- I.} Two emission lines at low energies were also needed. The lines are located at $E=0.56\pm0.01\rm\,keV$ ($EW=107^{+12}_{-27}\rm\,eV$) and $E=0.90\pm0.02\rm\,keV$ ($EW=68^{+37}_{-33}\rm\,eV$), and are consistent with being due to O\,VII and Ne\,IX, respectively. 
\smallskip\newline
{\bf PG 1126$-$041 -- II.} Two emission lines at $E=0.60\pm0.03\rm\,keV$ ($EW=70^{+6}_{-38}\rm\,eV$) and $E=0.89\pm0.03\rm\,keV$ ($EW\leq 280\rm\,eV$) are also needed. The lines are consistent with being due to O\,VII and Ne\,IX, respectively. 
\smallskip\newline
{\bf PG 1126$-$041 -- III.} The spectrum shows an additional emission line at $E=0.60\pm0.02\rm\,keV$ ($EW=115^{+53}_{-63}\rm\,eV$), consistent with the O\,VII line.
 \smallskip\newline
{\bf PG 1126$-$041 -- IV.} Two lines at low energy were also found. The line at $E=0.58\pm0.01\rm\,keV$ (EW=$65^{+12}_{-9}\rm\,eV$), is likely due to O\,VII emission, while that at $E=0.93\pm0.02\rm\,keV$ ($EW=89^{+22}_{-31}\rm\,eV$) is consistent with being Ne\,IX. Another emission line at $E=7.82_{-0.08}^{+0.13}\rm\,keV$ (EW=$87^{+24}_{-40}\rm\,eV$), consistent with being the He\,$\beta$ form of Fe\,XXV, was also found. 
 \smallskip\newline
{\bf PG 1229+204.} An emission line at $E=6.72\pm 0.05$ keV ($EW=98_{-25}^{+28}\rm\,eV$), consistent with being due to the He\,$\alpha$ state of Fe\,XXV, is required. 
\smallskip\newline
{\bf PG 1440+356 -- II .} An emission line at $E=6.72\pm 0.09$ keV ($EW=67_{-36}^{+19}\rm\,eV$) was detected, and is likely the He\,$\alpha$ form of Fe\,XXV. 
\smallskip\newline
{\bf PG 1440+356 -- III.} The spectrum shows evidence of emission due to the He\,$\alpha$ form of Fe\,XXV at $E=6.73\pm 0.10$ keV ($EW=106\pm41\rm\,eV$).
 \smallskip\newline
{\bf PG 1440+356 -- IV.}  We found evidence of an emission feature at $E=6.79\pm 0.07$\,keV ($EW=197\pm48\rm\,eV$), which is probably due to the He$\alpha$ state of Fe\,XXV. 
\smallskip\newline

 \bibliographystyle{aa}
 \bibliography{iron_cf}

\end{document}